\documentclass[amsfonts,amssymb,amsmath,aps,prb,notitlepage,superscriptaddress,longbibliography]{revtex4-2}%

\usepackage[T1]{fontenc}
\usepackage{times}

\usepackage{graphicx}
\usepackage{float}
\usepackage{soul}
\usepackage{dcolumn}
\usepackage{bm,color}
\usepackage{amsmath,amssymb,dsfont,amstext,amsfonts}
\usepackage{extarrows}
\usepackage[colorlinks=true,linkcolor=blue,urlcolor=blue,citecolor=blue]{hyperref}
\usepackage{xcolor}
\usepackage{wasysym}
\usepackage{mathtools}
\usepackage{bbold}
\usepackage{tikz}
\usepackage{orcidlink} 

\usepackage{graphicx}
\graphicspath{ {./images/} }

\usepackage[normalem]{ulem} 
\usepackage{color}

\newcommand{\hmbar}[1]{\ensuremath{\overline{\mkern-1mu#1\mkern-1mu}}}

\newcommand{\hmspace}[1]{\ensuremath{\mkern1mu#1\mkern1mu}}

\begin{document}
\title{Ultra-high spin emission from antiferromagnetic FeRh}

\author{Dominik Hamara\,\orcidlink{0000-0001-7318-8161}}
\affiliation{Microelectronics Group, Cavendish Laboratory, University of Cambridge, J.J.Thomson Avenue, Cambridge CB3 0HE, United Kingdom}

\author{Mara Strungaru\,\orcidlink{0000-0003-4606-7131}}
\affiliation{School of Physics, Engineering and Technology, University of York, York, YO10 5DD, United Kingdom}

\author{Jamie Massey\,\orcidlink{0000-0002-7793-7008}}
\affiliation{School of Physics and Astronomy, University of Leeds, Leeds LS2 9JT, United Kingdom}
\affiliation{Laboratory for Mesoscopic Systems, Department of Materials, ETH Zurich, 8093 Zurich, Switzerland.}
\affiliation{Paul Scherrer Institute, 5232 Villigen PSI, Switzerland.}

\author{Quentin Remy\,\orcidlink{0000-0002-4801-3199}}
\affiliation{Department of Physics, Freie Universit{\"a}t Berlin, 14195 Berlin, Germany}

\author{Guillermo Nava Antonio\,\orcidlink{0000-0003-2813-2841}}
\affiliation{Microelectronics Group, Cavendish Laboratory, University of Cambridge, J.J.Thomson Avenue, Cambridge CB3 0HE, United Kingdom}

\author{Obed Alves Santos\,\orcidlink{0000-0002-0192-8236}}
\affiliation{Microelectronics Group, Cavendish Laboratory, University of Cambridge, J.J.Thomson Avenue, Cambridge CB3 0HE, United Kingdom}

\author{Michel Hehn\,\orcidlink{0000-0002-4240-5925}}
\affiliation{Universit{\'e} de Lorraine, CNRS, IJL, F-54000 Nancy, France}

\author{Richard F.L. Evans\,\orcidlink{0000-0002-2378-8203}}
\affiliation{School of Physics, Engineering and Technology, University of York, York, YO10 5DD, United Kingdom}

\author{Roy W. Chantrell\,\orcidlink{0000-0001-5410-5615}}
\affiliation{School of Physics, Engineering and Technology, University of York, York, YO10 5DD, United Kingdom}

\author{St{\'e}phane Mangin\,\orcidlink{0000-0001-6046-0437}}
\affiliation{Institut Jean Lamour (UMR 7198), Universit{\'e} de Lorraine, Vandoeuvre-l{\`e}s-Nancy, France}

\author{Christopher H. Marrows\,\orcidlink{0000-0003-4812-6393}}
\affiliation{School of Physics and Astronomy, University of Leeds, Leeds LS2 9JT, United Kingdom}

\author{Joseph Barker\,\orcidlink{0000-0003-4843-5516}}
\email{j.barker@leeds.ac.uk}
\affiliation{School of Physics and Astronomy, University of Leeds, Leeds LS2 9JT, United Kingdom}

\author{Chiara Ciccarelli\,\orcidlink{0000-0003-2299-3704}}
\email{cc538@cam.ac.uk}
\affiliation{Microelectronics Group, Cavendish Laboratory, University of Cambridge, J.J.Thomson Avenue, Cambridge CB3 0HE, United Kingdom}

\begin{abstract}

An antiferromagnet emits spin currents when time-reversal symmetry is broken. This is typically achieved by applying an external magnetic field below and above the spin-flop transition \cite{Kholid_NatCommun_14_538_2023, Seki_PhysRevLett_115_266601_2015, Wu_PhysRevLett_116_097204_2016} 
or by optical pumping \cite{Qiu_NatPhys_17_388_2020}.
In this work we apply optical pump-THz emission spectroscopy to study picosecond spin pumping from metallic FeRh as a function of temperature. 
Intriguingly we find that in the low-temperature antiferromagnetic phase the laser pulse induces a large and coherent spin pumping, while not crossing into the ferromagnetic phase. 
With temperature and magnetic field dependent measurements combined with atomistic spin dynamics simulations we show that the antiferromagnetic spin-lattice is destabilised by the combined action of optical pumping and picosecond spin-biasing by the conduction electron population, which results in spin accumulation. We propose that the amplitude of the effect is inherent to the nature of FeRh, particularly the Rh atoms and their high spin susceptibility. We believe that the principles shown here could be used to produce more effective spin current emitters. Our results also corroborate the work of others showing that the magnetic phase transition begins on a very fast picosecond timescale\cite{Pressacco_NatCommun_12_5088_2021},
but this timescale is often hidden by measurements which are confounded by the slower domain dynamics.

\end{abstract}

\maketitle



\newpage

\section{Introduction}\label{sec:introduction}
{
FeRh has long been studied due to its first-order phase transition from a collinear antiferromagnet (AF) to a ferromagnet (FM) near room temperature \cite{Lewis_JPhysDApplPhys_49_323002_2016}. There is technological interest in a material that undergoes a significant change in its structural, magnetic, electrical, and thermal properties with just a small change in temperature \cite{Kryder_ProcIEEE_96_1810_2008,Marti_NatMater_13_367_2014} or strain \cite{Feng_AdvElectronMater_5_1800466_2018,Cherifi_NatMater_13_345_2014}. FeRh has also opened up a rich playground of fundamental studies to understand the nature of the phase transition and the parameters that trigger it. These studies have brought to light an intricate picture where the spin, electronic, and lattice degrees of freedom intertwine. 

The timescale of the phase transition triggered by laser pumping has been probed using techniques such as time-resolved X-ray diffraction and absorption spectroscopies \cite{Stamm_PhysRevB_77_184401_2008, Radu_PhysRevB_81_104415_2010, Mariager_PhysRevLett_108_087201_2012, bargheer2023}, magneto-optical Kerr effect \cite{Ju_PhysRevLett_93_197403_2004,Thiele_ApplPhysLett_85_2857_2004} and double pump-THz emission spectroscopy \cite{Li_NatCommun_13_2998_2022}. These studies showed that the ferromagnetic domain nucleation time is in the range of tens of picoseconds. Long-range ferromagnetic order in the direction of an external magnetic field follows, with typical timescales in the range of hundreds of picoseconds, limiting the speed of the magnetic phase transition. This picture naturally leads to consideration of the possibility of reducing the overall timescale of the phase transition by spin-biasing the FeRh at timescales shorter than the lattice response time. In this way, long-range magnetism would be established within the lattice response time, without the need for the magnetisation of the different domains to go through a slower realignment process.

In this work, we show that the free electron bath has a strong spin-polarising action on the antiferromagnetic phase in FeRh after optical pumping. A non-zero spin polarization in the free electron bath combined with optical pumping destabilises the antiferromagnetic order of the spin lattice and results in an enhanced generation of spin current. 

Optical pump-THz emission spectroscopy is a particularly suitable technique to measure optically generated spin currents at sub-picosecond timescales. Recently, this technique has been used in several studies of FeRh-Pt heterostructures at temperatures around the AF-FM phase transition, $T_\mathrm{(AF-FM)}\approx$ $350-370$~K \cite{Seifert_SPIN_07_1740010_2017, Medapalli_ApplPhysLett_117_142406_2020}. A femtosecond optical pump is used to heat the electron bath above ambient temperature. In the FM phase, this leads to a net spin transfer to the adjacent Pt layer, where the spin current is converted into a fast current pulse via the inverse spin Hall effect (ISHE) and results in the emission of THz electrodipole radiation. In the AF phase, the THz emission is negligible and is understood from the spin-degeneracy of the electron bands meaning there is zero net spin transfer to the Pt.
In this work, we have extended these studies over a wider temperature range of 20–420~K. We observe a previously unreported increase in the THz emission at low temperatures, consistent with a large spin current, despite the negligible magnetisation in the AF phase. This cannot be explained within a standard spintronic emitter picture, and our work provides new insight into exchange-enhanced spin current generation in FeRh mediated by the Rh spin.}

\newpage

\section{Experimental Results}
\subsection{Optical Pump-THz emission from FeRh-Pt}
In our measurements we used two thin-film structures: MgO/FeRh(Pd,Ir)(30)/Pt(3.5) and MgO/FeRhPd(35), with layer thicknesses given in nanometres, respectivelly referred to as FeRh/Pt and FeRh samples. The composition of the FeRh/Pt sample is MgO/Fe$_{\text{50}}$Rh$_{\text{46.8}}$Pd$_{\text{1.7}}$Ir$_{\text{1.5}}$(5)/Fe$_{\text{50}}$Rh$_{\text{47.1}}$Pd$_{\text{2.2}}$Ir$_{\text{0.7}}$(10)/Fe$_{\text{50}}$Rh$_{\text{47.2}}$Pd$_{\text{2.8}}$(15)/Pt(3.5).  The Pd and Ir doping gradients result in a transition temperature gradient \cite{Barua_ApplPhysLett_103_102407_2013, Temple_ApplPhysLett_118_122403_2021}. The FeRh sample is a uniform Fe$_{\text{50}}$Rh$_{\text{47.2}}$Pd$_{\text{2.8}}$ film used as a control.

\begin{figure}[h]
    \centering
\includegraphics[width=0.8\textwidth]{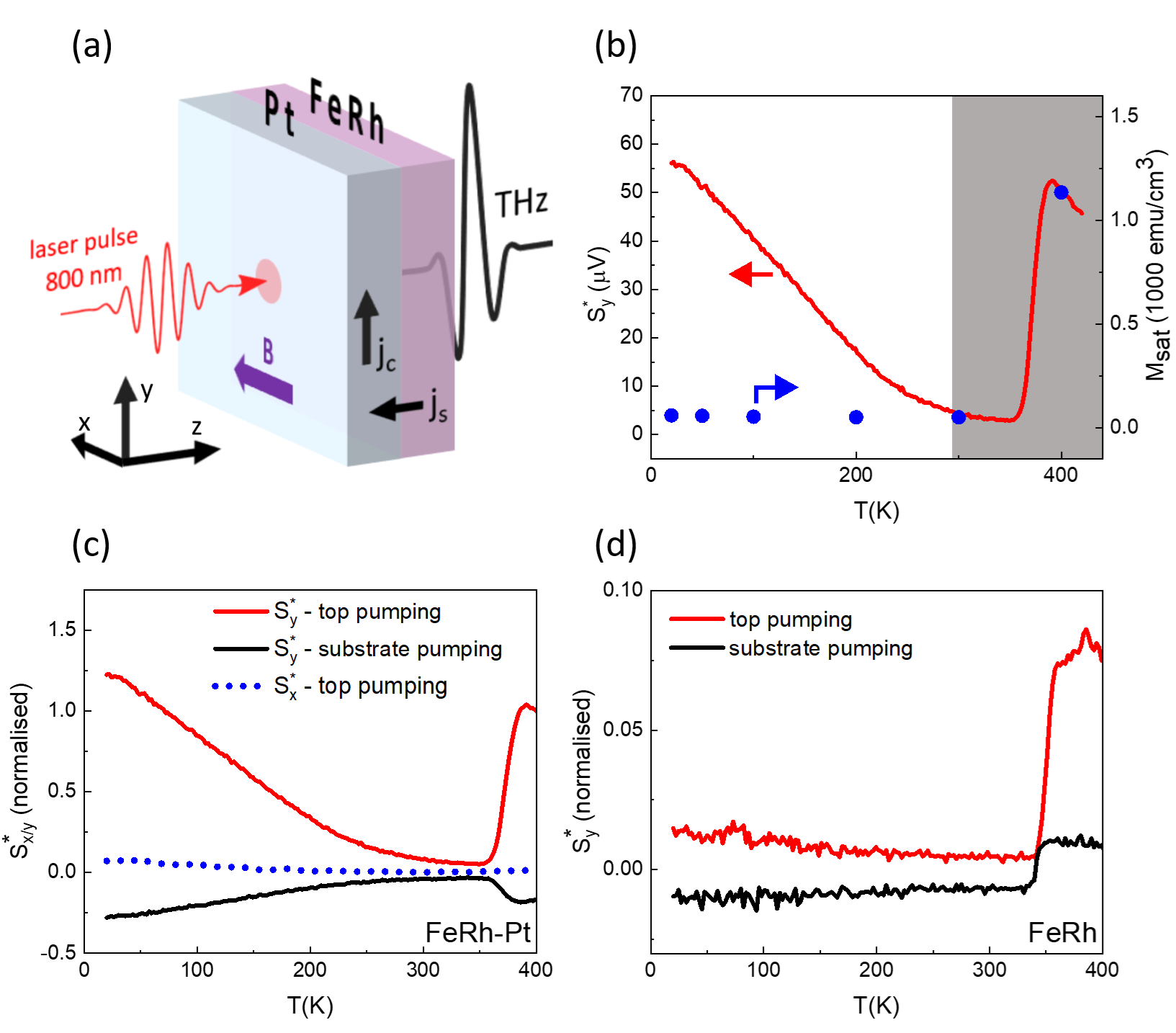}
\caption{ 
    \textbf{(a)} Layout of the experiment. The pump pulse propagates in the $z$ direction, while a magnetic field can be applied in the $x$ direction. $j_s$ represents the spin current, while $j_c$ is the charge current generated in the Pt layer via the ISHE. \textbf{(b)} Temperature dependence of the THz emission amplitude, rescaled by the optical absorption and THz outcoupling, $S_y^*$ (red line), in top-pumping geometry with an applied magnetic field of 860 mT. On the same graph, we plot the saturation magnetisation $M_\mathrm{sat}$ as a function of temperature (blue dots), derived from isothermal hysteresis loops shown in section S2 A. of the SI. \textbf{(c)} Temperature dependence of $S_y^*$ for two different pumping geometries - top pumping (red) and substrate pumping (black) - and an applied magnetic field of 860 mT for the FeRh-Pt sample. The blue dotted line shows the temperature dependence of the rescaled THz emission amplitude polarised along the $x$ axis, $S_x^*$, for a top-pumping geometry and the same value of the applied magnetic field. \textbf{(d)} Temperature dependence of $S_y^*$ for two different pumping geometries - top pumping (red) and substrate pumping (black) - and an applied magnetic field of 860 mT for the uncapped FeRh sample. In this case the normalisation factor used to extract $S_y^*$ was calculated using the electrical conductivity of the uncapped FeRh sample. Data in both \textbf{(c)} and \textbf{(d)} are normalised by the value of $S_y^*$(400 K) measured in FeRh-Pt in the top-pumping geometry. Data sets without the correction for THz outcoupling and information on the absorbed pump fluences are included in section S1 of the SI.}
    \label{fig:fig1}
\end{figure}

The samples were grown using DC magnetron sputtering on $(001)$-oriented MgO substrates. The substrates were annealed \textit{in situ} overnight at 600$^\circ$~C. After the annealing process, the FeRh thin-film was deposited. The sample was then annealed \textit{in situ} at 700$^\circ$~C for 1 hour. The samples were transferred to a different deposition chamber to deposit the Pt. In this chamber, the samples were reannealed at 700$^\circ$~C for an hour and a half to refine the crystal structure and remove any defects on the surface that formed between the depositions. The sample was then left to cool to room temperature before the Pt layer was deposited.

The layout of the experiment is shown in Fig.~\ref{fig:fig1}(a). We applied femtosecond laser pulses incident on the sample and measured the emitted THz electric field pulses in the time-domain by electro-optic sampling with a 1~mm ZnTe crystal. To indicate different directions we introduce a Cartesian coordinate system fixed with respect to our setup, where the $z$-axis is normal to the sample surface. Pump pulses propagate along the $z$-axis. The sample was mounted in an optical cryostat and the ambient temperature was varied in the range $20-420$~K. We applied an external magnetic field along the $x$-axis using an electromagnet, with a field strength up to $|B|\approx 860$~ mT.  We pumped the sample from the substrate side (substrate pumping) or the capping layer side (top pumping). For the optical excitation we used linearly-polarised 50~fs laser pulses with central wavelength of 800~nm. 

The THz radiation emitted after optical pumping has mainly two origins: i) the magneto-dipole component, which is directly related to the pump-induced change in magnetisation ($\frac{dM}{dt}$), and ii) the electro-dipole component, which is related to the spin current generated, $j_\mathrm{s}$, converted into a radiating sub-picosecond charge current via the ISHE. The detected far-field THz radiation, $S(\omega)$, is
related to the THz field directly behind the sample, $E_\mathrm{THz-out}$, by the convolution
with the transfer function of the detection unit \cite{Kampfrath2018}. In the case of a purely electro-dipole emission \cite{Seifert_SPIN_07_1740010_2017}:
\begin{equation}
\label{eqn:Eq1}
   S(\omega)\propto  E_\mathrm{THz-out} (\omega) \propto C(\omega)   E_\mathrm{THz-in} (\omega)\propto A \theta_{\mathrm{SH}} \lambda_\mathrm{s} C(\omega)  j_\mathrm{s} (\omega).
\end{equation}
Here, $E_\mathrm{THz-in}(\omega)$ represents the THz electric field generated within the sample, $A$ is the optical fluence absorbed by the metallic stack at the pump wavelength of 800 nm, $\theta_{\mathrm{SH}}$ is the spin Hall angle and $\lambda_\mathrm{s}$ is the spin relaxation length. The parameter $C(\omega)$ describes the outcoupling of the THz field from the metallic stack:
\begin{equation}
\label{eqn:EqS4}
   C(\omega)\propto\frac{1}{1+n_\mathrm{MgO}+Z_0\int_{0}^{d}{dz\ \sigma(z)}} .
\end{equation}
where $n_\mathrm{MgO}$ is the refractive index of the MgO substrate, $Z_0$ is the vacuum impedance, $d$ is the metal stack thickness, and $\sigma(z)$ is the conductivity distribution of the metal stack.

\subsection{Temperature dependence}

In Fig.~\ref{fig:fig1}(b) we plot $S^* _y (T) = S_y(T)/[C(T) A(T)]$, the peak-to-peak amplitude of the emitted THz pulse transient normalised by the temperature-dependent THz outcoupling $C(T)$ and optical pump absorption $A(T)$, as described in detail in section S1 of the Supplementary Information (SI). The subscript $y$ indicates the pulse component polarised along the $y$ axis, perpendicular to the applied magnetic field. Normalising by $C(T)$ and $A(T)$ accounts for the temperature dependence of the sample's conductivity and refractive indices, therefore in the case of electro-dipole emission, the temperature dependence of $S^* _y$ essentially reflects the temperature dependence of the spin current, $j_\mathrm{s}$.

Previous optical pump-THz emission studies on FeRh and FeRh-Pt \cite{Seifert_SPIN_07_1740010_2017, Medapalli_ApplPhysLett_117_142406_2020, Awari_ApplPhysLett_117_122407_2020},  focused on the metamagnetic AF-FM phase transition near $T_\mathrm{(AF-FM)}$ and only measured down to $T=300$~K. The shaded area in Fig.~\ref{fig:fig1}(b) indicates the temperature region studied in those works. In this range, we observe the same behaviour as others, with the THz emission amplitude being almost fully suppressed below $T_\mathrm{(AF-FM)}$ due to the reduction in volume of the FM phase. However, we extended the measurements by continuing to the low temperature region, down to $T=$~20~K. This is deep within the AF phase and far below $T_\mathrm{(AF-FM)}$ which is 370~K in our sample for the absorbed pump fluence of 2.4~mJ/cm$^2$. Surprisingly, as we decrease the temperature, $S^* _y (T)$ starts to increase again, above the value in the FM phase. This is despite the fact that there is only a tiny magnetisation in the AF phase, which we measured with SQUID (Superconducting Quantum Interference Device). Full data is shown in section S2 A. of the SI. 

Below $T_\mathrm{(AF-FM)}$ the THz emission is linear with the pump fluence (see the section S3 E. of the SI) and we see no threshold behaviour. These observations indicate that the increasing $S^*_y$ in the AF phase is not related to a heat-induced metamagnetic phase transition. From the specific heat of FeRh (0.35 J/g K) \cite{PhysRevMaterials.5.064412} we estimate that the transient electron temperature rise upon optical pumping is $\sim$ 130 K in the first 10 nm near the Pt interface at the pump fluence of 2.4 mJ/cm$^2$, too low to justify a phase transition at low temperature. Moreover, heat accumulation effects are minimal due to the low repetition rate of our laser (5 kHz), as also discussed in section S3 G. of the SI.

We confirm that the THz emission is of electro-dipole character, induced by a spin current, by measuring in both the top and substrate pumping geometries. The results in Fig.~\ref{fig:fig1}(c) show that the THz emission has a dominant odd component when the sample is flipped with respect to the pump incidence side, in agreement with the symmetry of the ISHE. We see the ISHE symmetry both above and below $T_\mathrm{(AF-FM)}$ which further confirms that in both the FM and AF phases the emission is due to spin currents. The smaller amplitude of the emission when pumping from the substrate side is because of the smaller absorbed pump energy in the region closer to the interface with Pt. On the same graph, we show that the THz emission is fully polarised along the $y$ direction, which is compatible with a polarisation of the spin currents parallel to the external field. 

In Fig.~\ref{fig:fig1}(d) we show the same top- and substrate-pumping measurements performed on an uncapped FeRh sample without any Pt. The temperature dependence of the THz emission differs significantly from that in FeRh-Pt. In the ferromagnetic phase, the amplitude of the THz emission is more than a factor of six smaller. Moreover, it does not show the expected symmetry for electro-dipole emission; its polarity does not change for two flipped pumping geometries. This agrees with previous observations \cite{Li_NatCommun_13_2998_2022} and the signal is attributed to the magneto-dipole origin of the emission, which can become the main contribution in absence of efficient spin-to-charge conversion via the ISHE. At $T < T_\mathrm{(AF-FM)}$ the emission is still smaller, but shows an increasing trend as the temperature is lowered. Quite remarkably, in this AF region, the flipped geometry changes sign as expected for spin currents converted via the ISHE. Bare FeRh does have an appreciable spin Hall angle and the quantity $\theta_{\mathrm{SH}}\lambda_s$ has been measured to be twice as large in the AF phase than the FM phase \cite{Wang_NatCommun_11_275_2020}. Even so, this result indicates that in the AF phase where negligible spin current would normally be expected, there is actually a bulk contribution large enough to produce an electro-dipole emission. We note that the AF phase of FeRh usually has some residual FM order \cite{Almeida_PhysRevMater_4_034410_2020, Song_ApplSurfSci_607_154870_2023}, often as a surface skin, but the contribution from these FM regions alone would be expected to be much smaller than the bulk FM phase.

\subsection{Magnetic field dependence}
\begin{figure}[h]
    \centering
\includegraphics[width=0.8\textwidth]{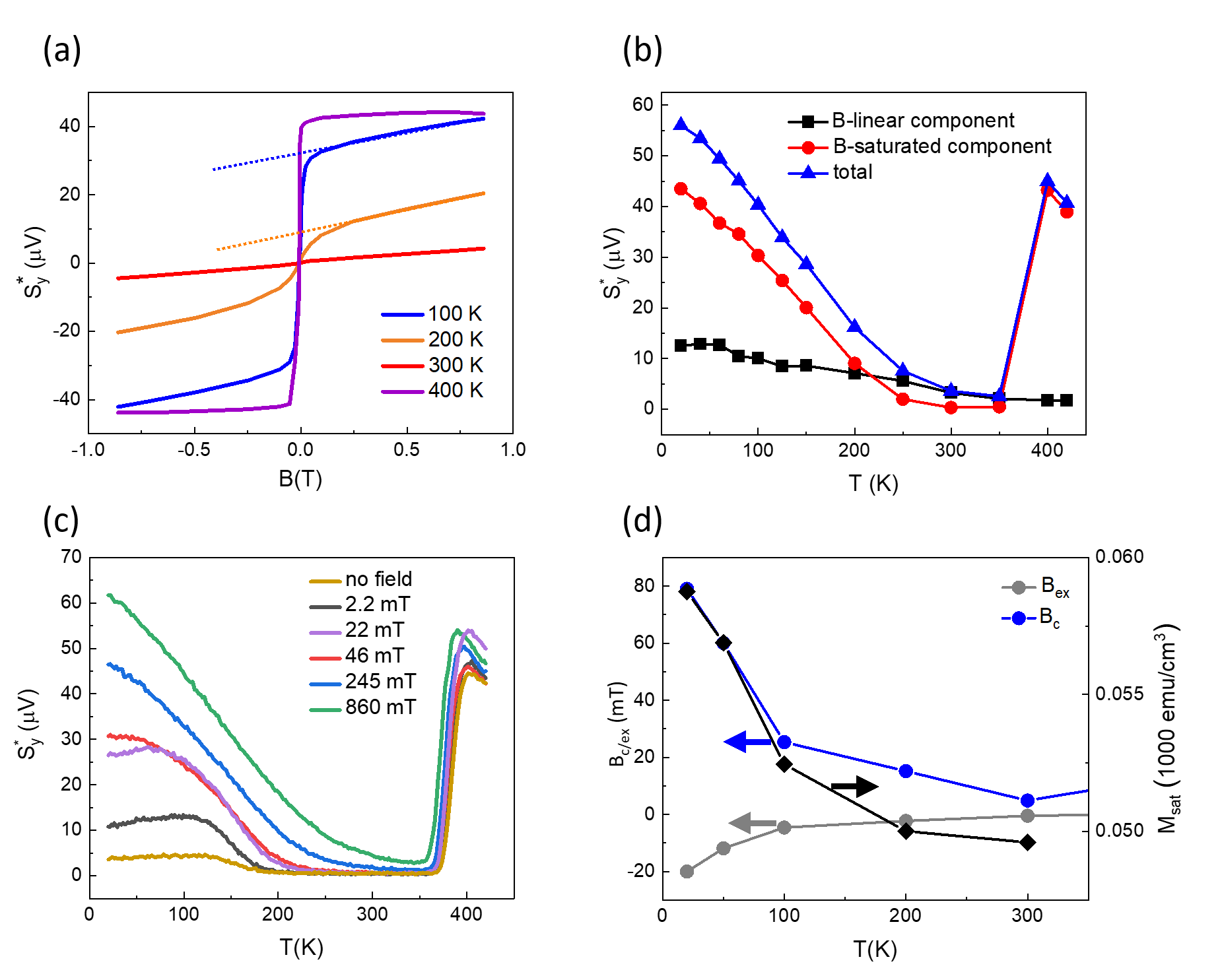}
\caption{\textbf{(a)} THz emission amplitude measured as a function of in-plane magnetic field at selected temperatures. The dashed lines indicate the B-linear components of the signals. \textbf{(b)} Amplitude of both components $s_{y(\mathrm{B-sat})}$ and $s_{y(\mathrm{B-linear})}$ and their sum at +860~mT over the investigated temperature range.   \textbf{(c)} Temperature dependence of $S_y^*$ for different magnetic fields up to +860~mT. The experiment was performed in the top-pumping geometry. Data sets without the correction for THz outcoupling and the information on the absorbed pump fluences are included in the SI. \textbf{(d)} Temperature dependence of $M_\mathrm{sat}$, coercive field $B_\mathrm{c}$ and exchange bias field $B_\mathrm{ex}$ of the residual FM regions in FeRh-Pt.}
    \label{fig:fig2}
\end{figure}

Fig.~\ref{fig:fig2}(a) shows the magnetic field dependence of $S_y^*$ in the FeRh-Pt sample at fixed temperatures. In the ferromagnetic phase ($T=400$~K), $S_y^*$ saturates with a small magnetic field, as we expect in a spintronic emitter picture where the pump-generated spin currents scale with the magnetisation of the ferromagnetic metal. In the antiferromagnetic phase ($T=100$~K, $T=200$~K) $S_y^*$ contains both a saturating behaviour and a linear increase. At a sufficiently high magnetic field we can describe the signal with $S_y^* = S^* _{y(\mathrm{B-sat})} + S^* _{y(\mathrm{B-linear})} B$. In Fig.~\ref{fig:fig2}(b) we plot $S^* _{y(\mathrm{B-sat})}$ and $S^* _{y(\mathrm{B-linear})}$ as a function of temperature with an applied field of +860 mT. In the considered field range $S^* _{y(\mathrm{B-sat})}$ has a significantly larger amplitude with respect to $S^* _{y(\mathrm{B-linear})}$ at low temperature. Also, differently from $S^* _{y(\mathrm{B-sat})}$ that keeps increasing with decreasing temperature, $S^* _{y(\mathrm{B-linear})}$ saturates at $T \approx 50$~K. The $S^* _{y(\mathrm{B-linear})}$ component is consistent with the Zeeman splitting of AF states \cite{Kholid_ApplPhysLett_119_032401_2021, Kholid_NatCommun_14_538_2023} allowing for non-zero net spin currents. The $S^* _{y(\mathrm{B-sat})}$ component cannot be easily understood based on previous works, and in the following discussion we focus solely on this component.

The magnetic field dependence of $S_y^*$ suggests that residual ferromagnetism in the FeRh-Pt bilayer plays a role in the spin current generation below $T_{(AF-FM)}$. The presence of ferromagnetism at both FeRh interfaces, with the substrate and the capping layer, is well known \cite{Fan_PhysRevB_82_184418_2010, Han_JApplPhys_113_17C107_2013, Kinane_NewJPhys_16_113073_2014, Zhou_AIPAdv_6_015211_2016, Gatel_NatCommun_8_15703_2017,Baldasseroni_ApplPhysLett_100_262401_2012, Ding_JApplPhys_103_07B515_2008, Song_ApplSurfSci_607_154870_2023}. Depth profile studies \cite{Fan_PhysRevB_82_184418_2010, Song_ApplSurfSci_607_154870_2023,Almeida_PhysRevMater_4_034410_2020} have shown that these FM regions typically extend over a thickness of 3-5~nm and have a much lower average magnetic moment relative to the FM phase of FeRh.
We confirm the presence of ferromagnetism in the low-temperature region in our samples using SQUID magnetrometry (section S2 A. of the SI). In Fig. \ref{fig:fig2}(d) we plot the saturation magnetisation $M_\mathrm{sat}$ extracted from the SQUID data below $T_{(AF-FM)}$ and show that it increases slightly as we decrease the temperature below $T \approx 200$~K. The data shown in  Fig.~\ref{fig:fig2}(d) is normalised by the total volume of FeRh, but even considering the normalisation by the effective volume of the ferromagnetic interfaces with a combined thickness of 10~nm, $M_\mathrm{sat}$ in the AF phase remains significantly smaller than $M_\mathrm{sat}$ for FM-FeRh, in agreement with previous studies \cite{Hashi_IEEETransMagn_40_2784_2004, Lewis_JPhysDApplPhys_49_323002_2016, Eggert_RSCAdv_10_14386_2020, Song_JAlloyCompd_901_163611_2022}. We reproduce the temperature dependence of the magnetisation via atomistic spin dynamics  simulations considering  a monolayer of FM-FePd in contact with AF-FeRh, as discussed in more detail in section S2 B. of the SI.

From the SQUID data we also extract a positive ($B_\mathrm{c+}$) and negative ($B_\mathrm{c-}$) coercive field value as a function of ambient temperature. Fig.~\ref{fig:fig2}(d) shows that the coercivity, defined as $B_\mathrm{c}=(B_\mathrm{c+}-B_\mathrm{c-})/2$, increases below 300~K, in line with other FM alloys \cite{Vertesy_JApplPhys_71_3462_1992, Vertesy_JApplPhys_77_6426_1995, Turtelli_PhysRevB_66_054441_2002}. At temperatures below 300~K we also observe a clear asymmetry between $B_\mathrm{c+}$ and $B_\mathrm{c-}$, indicated by a non-zero value of $B_\mathrm{ex}=B_\mathrm{c+}+B_\mathrm{c-}$. This is likely due to an exchange interaction between the FM interface and the adjacent antiferromagnet. 

In Fig. \ref{fig:fig2}(c) we plot the temperature dependence of $S^* _y$ at different magnetic field values. Here we observe that while at the highest field values above $B_c$  (860 mT and 245 mT) the curves are described by the same temperature law, apart from a linear-in-field offset, at lower field values below $B_\mathrm{c}$,  $S_y^*$ is heavily suppressed, pointing again towards a role of the FM interface. 

\section{Discussion}

\begin{figure}[h]
    \centering
\includegraphics[width=0.8\textwidth]{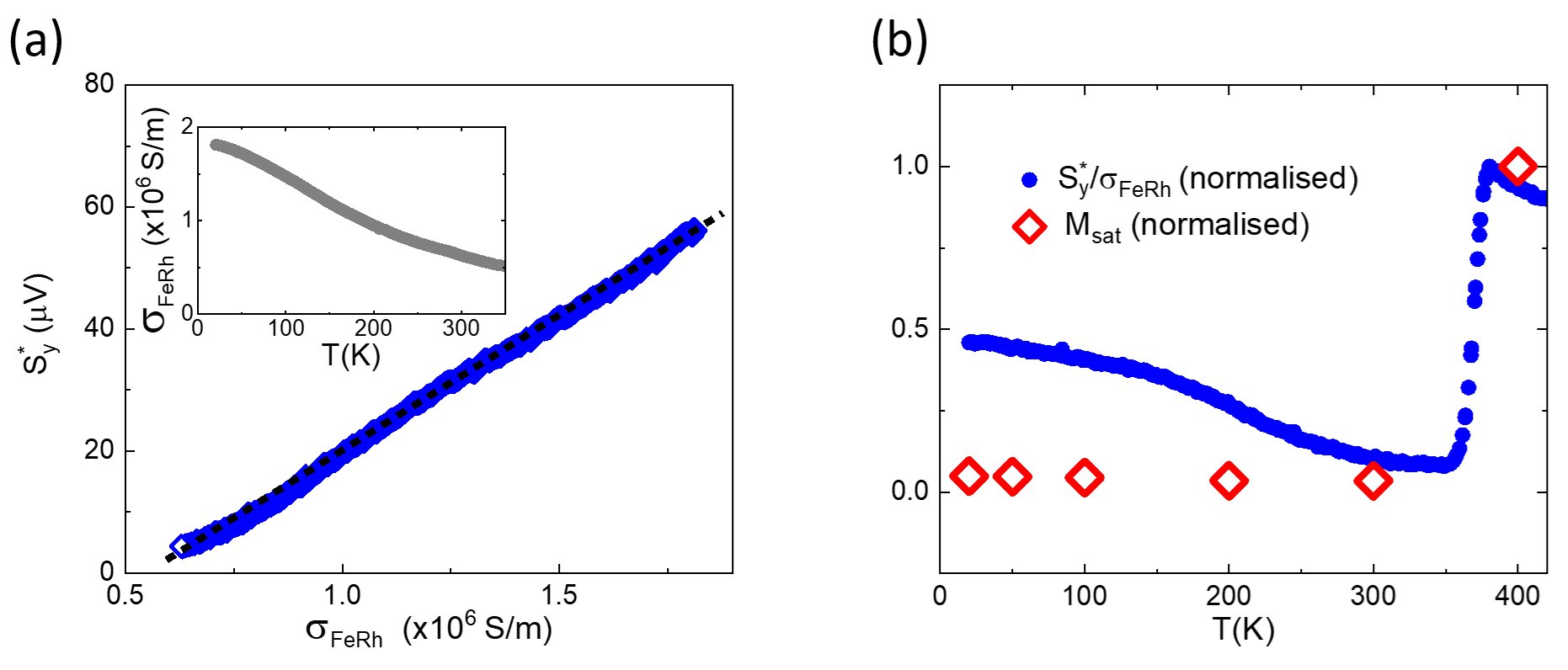}
\caption{\textbf{(a)} $S_y^*$ measured in FeRh-Pt, pumping on the Pt side, as a function of FeRh conductivity, $\sigma_\mathrm{FeRh}$ and for a value of the external field of 860 mT. The insert shows the temperature dependence of $\sigma_\mathrm{FeRh}$ \cite{Almeida_PhysRevMater_4_034410_2020}. \textbf{(b)} Temperature dependence of normalised $S^*_y/ \sigma_\mathrm{FePt}$ at an applied field of 860 mT (blue dots) and of normalised $M_\mathrm{sat}$ (red diamonds).}
    \label{fig:fig3}
\end{figure}

The temperature and magnetic field dependence of $S^* _y$ suggests that the residual magnetism, likely to exist at the interface with Pt, plays a key role in the enhanced spin emission at low temperature. However, the magnitude of $S^* _y$ in the AF phase cannot be explained by contributions from such a small magnetisation and small volume of the FeRh if only the small FM regions contribute to the spin currents. Therefore, we conclude that the antiferromagnetic phase must be contributing a significant spin current. For the antiferromagnet to act as a source of spin angular momentum time-reversal symmetry must be broken. The Zeeman splitting in the applied field provides this in a small way, and we have already identified this contribution as the linear term $S^* _{y(\mathrm{B-linear})}$. Optical pumping by itself does not break time-reversal symmetry. The magnetic difference frequency generation process \cite{Higuchi_PhysRevLett_106_047401_2011, Qiu_NatPhys_17_388_2020} does not apply here because we pump along the (001) crystal axis which is not a three-fold rotation axis (FeRh belongs to space group \hmspace{P}\hmspace{m}\hmbar{3}\hmspace{m}). The optical pumping therefore only results in the diffusion of spin polarised hot carriers from the FM volume. In the standard spintronic emitter picture, only the transport of these carriers towards Pt is considered. We believe that transport towards the AF bulk of FeRh is instead important to consider in order to explain our results, as suggested by the proportionality of $S^*_y$ with the FeRh DC conductivity, $\sigma_\mathrm{FeRh}$, below $T_\mathrm{(AF-FM)}$ in Fig. \ref{fig:fig3}(a). In a spintronic emitter picture, instead, the temperature dependence of the spin current is dictated by the spin Hall physics of Pt and scales with its resistivity $\rho_\mathrm{Pt} (T)$ \cite{Matthiesen_ApplPhysLett_116_212405_2020}, please refer to section S3 H of the SI. Our observation suggests that charge transport in FeRh is important to the generation of the spin currents at low temperature. In Fig. \ref{fig:fig3}(b) we plot the temperature dependence of $S^*_y/ \sigma_\mathrm{FePt}$. In the region of the phase transition, $S^*_y/ \sigma_\mathrm{FePt}$ scales with the drop in $M_\mathrm{sat}$, as one would expect in a spintronic emitter picture. The significantly higher rate of increase as temperature is further lowered below $T_{(AF-FM)}$ reflects the different origin of the spin current in the AF phase.

We therefore suggest that the spin-polarised free carriers from the interfacial region have a strong polarising action on the FeRh lattice after optical pumping. Time-resolved photoelectron spectroscopy studies have shown that optical pumping of AF-FeRh results in a sub-picosecond alteration of the electronic structure that redistributes spin density from Fe towards Rh, producing a small transient Rh moment \cite{Pressacco_NatCommun_12_5088_2021}. This is a manifestation of the optically induced intersite spin transfer (OISTR) process~\cite{Elliott_SciRep_6_38911_2016, Dewhurst_NanoLett_18_1842_2018, Hofherr_SciAdv_6_eaay8717_2020}. 
Despite this redistribution of spin density, unless the system is macroscopically spin-biased, the transient Rh moments in the different unit cells will average to zero and will not result in any coherent action on the Fe spin-lattice. 

\begin{figure}[h]
    \centering
\includegraphics[width=0.8\textwidth]{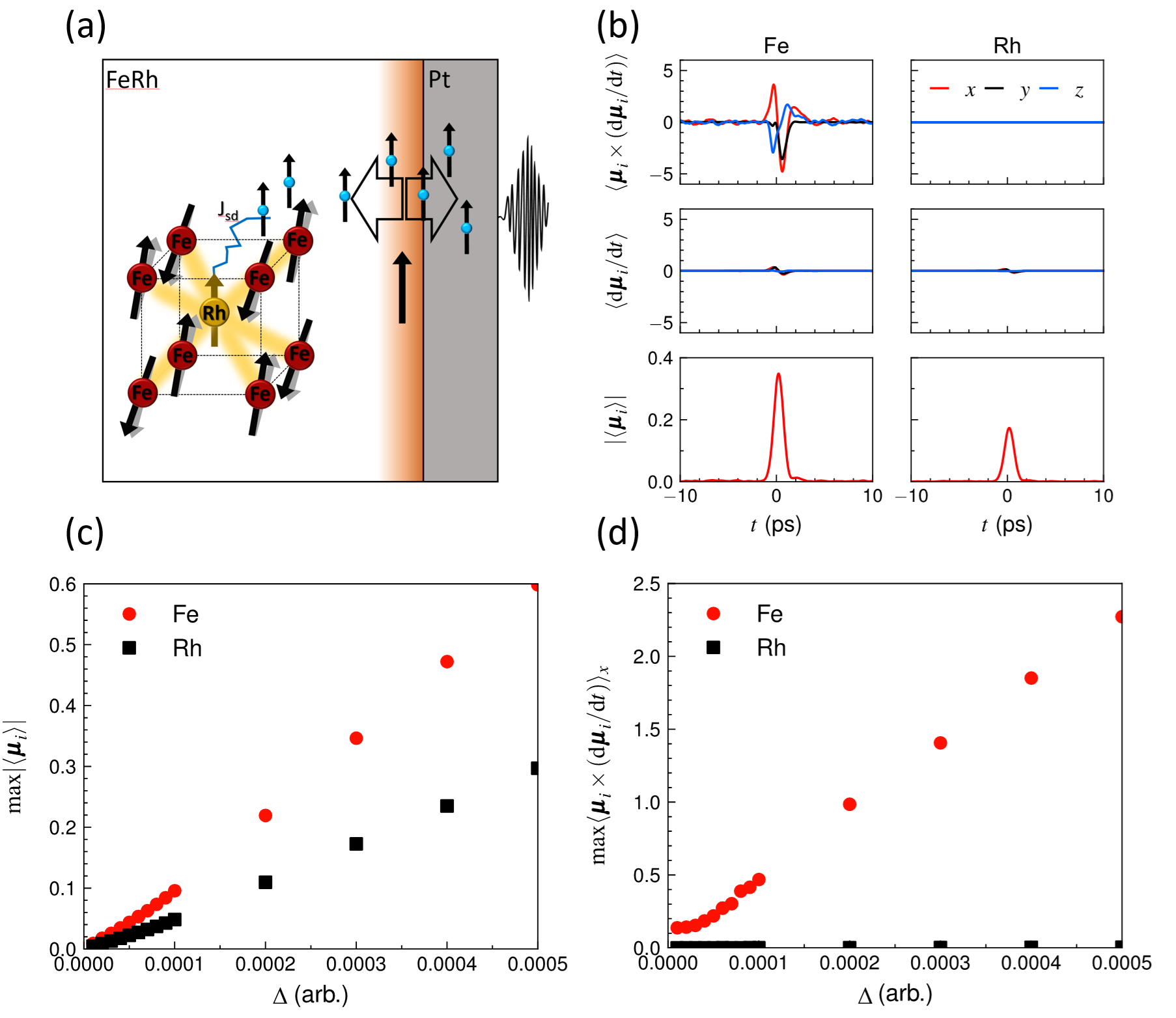}
\caption{\textbf{(a)}  Schematics of the mechanism that leads to spin pumping in the AF phase of FeRh. Optical pumping causes the diffusion of spin-polarised carriers from the ferromagnetic interface towards AF-FeRh, inducing a transient spin on the Rh atoms (shown by the brown arrow). The Rh spin acts with a torque on the Fe spins, shown by black arrows, and the magnetisation rises. The generated non-equilibrium spin in FeRh is pumped back into Pt and results in THz emission (not shown). \textbf{(b)}  Dynamics of real (top) and imaginary (middle) parts of the spin pumping and the size of the induced magnetisation (bottom) in the Fe (left) and Rh (right) sublattices with $\Delta=3\times10^{-5}$. \textbf{(c)}  Maximum amplitude of the induced magnetisation in Fe (red circles) and Rh (black squares) as a function of the pulse amplitude ($\Delta$). \textbf{(d)}  Size of the maximum in the real spin pumping along the $x$ (applied field) direction as a function of the pulse amplitude ($\Delta$).}

    \label{fig:simulations}
\end{figure}

The spin-polarised free carriers from the ferromagnetic interface provide this spin-bias via s-d exchange, with their polarising action increasing with the spin lifetime within FeRh, as schematically shown in Fig. \ref{fig:simulations}(a). A recent work by Kang and co-workers \cite{Kang_NatCommun_14_3619_2023} showed that the optically-induced AF-to-FM phase transition results in spin angular momentum absorption by the conduction electron bath of a Cu capping layer. In this work, we show instead that spin-polarised conduction electrons strongly destabilise the AF lattice and spin polarise it also at temperatures much below the phase transition. To study the effect of a small, polarised, but transient moment being induced on the Rh sites, we performed atomistic spin dynamics simulations including a Landau Hamiltonian term which allows the magnitude of the spin moments to change and fluctuate~\cite{Ma_PhysRevB_86_054416_2012, Derlet_PhysRevB_85_174431_2012, Sandratskii_PhysRevB_83_174408_2011}. We model the polarising effect as a Gaussian increase in the size of the Rh moments, in the spin polarised direction, while the laser pulse is applied. A full description of the model is in section S4 of the SI.

We perform spin dynamics simulation where we apply a weak perturbation that consists in a non-zero transient spin on the Rh site, less than 0.2~$\mu_{\mathrm{B}}$ per Rh spin on average (Fig.~\ref{fig:simulations}(b), bottom right), and find a significant coherent spin pumping from the Fe spin-lattice (Fig.~\ref{fig:simulations}(b), top left). The sudden appearence of the Rh moment within the lattice breaks the time-reversal symmetry. The strong Fe-Rh exchange amplifies the effect of the Rh and results in a net magnetisation of $\sim$0.35~$\mu_{\mathrm{B}}$ per Fe spin (Fig.~\ref{fig:simulations}(b), bottom left) which is much larger that the magnetisation induced by Zeeman splitting for the magnetic fields we apply. The Fe moments precess in the exchange field from the Rh, producing a spin pumping (Fig.~\ref{fig:simulations}(b), top row). To generate a net Fe moment, the exchange field must be greater than the spin flop field of the AF phase which we estimate to be 30~T. For an induced Rh moment of 0.2~$\mu_{\mathrm{B}}$ the exchange field at the peak is around 200~T, although appearing for less than a picosecond (see section S5 of the Supplementary Information (SI) for more details). We note that the longitudinal spin pumping from both the Fe and Rh represents a negligible contribution (Fig.~\ref{fig:simulations}(b), middle row). In fact, the Rh moments contribute almost no direct spin pumping, differing from predictions at the phase transition \cite{Kang_NatCommun_14_3619_2023}, but their effect of breaking the time-reversal symmetry on the Fe spin-lattice is dramatic. This results in spin angular momentum being released from the antiferromagnetic phase, even without causing the metamagnetic phase transition. Increasing the applied magnetic field results in increased ordering of the ferromagnetic skin, until the magnetisation saturation point. The increasing magnetic order results in the higher polarisation of the initial spin current, leading to the increase in $S_y^*$, as we observe in our experiment. We simulate this by studying how the induced magnetisation (Fig.~\ref{fig:simulations}(c)) and spin pumping (Fig.~\ref{fig:simulations}(d)) increase with $\Delta$, the amplitude of the transient Rh moment. We find that the induced magnetisation of both Fe and Rh are linear with $\Delta$. For the spin pumping, we see that it is dominated by the Fe contribution and increases with $\Delta$, in agreement with the experiments where larger magnetic fields (more strongly ordering the FM skin) and higher laser fluence both result in increases in the spin current. In the experiments, the Rh-mediated exchange-enhanced spin pumping saturates above the coercive field of the residual ferromagnetic interface ($\sim$ 80 mT), because the spin polarisation of the currents generated by optical pumping saturates. The spin pumping that we measure exceeds that measured in ferromagnetic metal CoFeB, as shown in section S3 H of the SI. We predict that the amplitude of the spin pumping could be further enhanced by doping engineering to control the position and nature of the ferromagnetic regions such to maximise the antiferromagnetic volume involved in the spin emissison. The possibility to generate high spin current pulses at low magnetic fields is important in the context of high-speed magnetic recording \cite{mangin20,mangin18}. Achieving this with an antiferromagnet would suppress dipolar interactions, allowing growing the spin emitter in direct contact with the active bit-element.



\section*{Data availability}
The datasets generated during and/or analysed during the current study will be made available in the Repository of the University of Cambridge Apollo at the address \url{https://doi.org/10.17863/CAM.107667}. The \textsc{vampire} software used for the four spin model is available at \url{https://vampire.york.ac.uk}. The \textsc{JAMS} software used for the induced moment simulations is not currently open source while intellectual property issues are being resolved, but can be made available to individual researchers upon request. 

\begin{acknowledgements}
CC and JB acknowledge support from the Royal Society through University Research Fellowships. This project was supported by the Diamond Light Source and has received funding from the European Union’s Horizon 2020 research and innovation programme under the Marie Sk\l{}odowska-Curie (grant agreement No. 861300) and the Engineering and Physical Sciences Research Council (grant numbers EP/V037935/1 and EP/X027074/1). Calculations were performed on ARC4, part of the High Performance Computing facilities at the University of Leeds. CC thanks Dr. Samer Kurdi for the fruitful discussion. QR would like to thank Karel V\'{y}born\'{y} for the fruitful discussion on ref. [4] of the SI as well as for providing data necessary to get the interband conductivity (ref. [5] in the SI). 
\end{acknowledgements}

\section*{Author Contributions}

\textbf{Dominik Hamara}: data curation, formal analysis, methodology, visualisation, writing-original draft.
\textbf{Mara Strungaru}: formal analysis, software, visualisation, writing–original draft.
\textbf{Jamie R. Massey}: methodology.
\textbf{Quentin Remy}: formal analysis, software.
\textbf{Guillermo Nava Antonio}: formal analysis, methodology, writing–review and editing.
\textbf{Obed Alves Santos}: methodology.
\textbf{Michel Hehn}: methodology.
\textbf{Richard F.L. Evans}: supervision, visualisation.
\textbf{Roy W. Chantrell}: formal analysis, software, supervision.
\textbf{St{\'e}phane Mangin}: formal analysis, methodology.
\textbf{Christopher H. Marrows}: methodology, supervision, funding acquisition.
\textbf{Joseph Barker}: formal analysis, funding acquisition, software, writing-original draft.
\textbf{Chiara Ciccarelli}: conceptualization, formal analysis, funding acquisition, project administration, supervision, writing-original draft.

\section*{Competing interests}
The authors declare no competing interests.

\newpage
\bibliography{references}


\end{document}


\title{Supplementary Information: Ultra-high Spin Emission from Antiferromagnetic Metal FeRh}
\author{Dominik Hamara}
\affiliation{Microelectronics Group, Cavendish Laboratory, University of Cambridge, J.J.Thomson Avenue, Cambridge CB3 0HE, United Kingdom}

\author{Mara Strungaru\,\orcidlink{0000-0003-4606-7131}}
\affiliation{School of Physics, Engineering and Technology, University of York, York, YO10 5DD, United Kingdom}

\author{Jamie R. Massey\,\orcidlink{0000-0002-7793-7008}}
\affiliation{School of Physics and Astronomy, University of Leeds, Leeds LS2 9JT, United Kingdom}
\affiliation{Laboratory for Mesoscopic Systems, Department of Materials, ETH Zurich, 8093 Zurich, Switzerland.}
\affiliation{Paul Scherrer Institute, 5232 Villigen PSI, Switzerland.}

\author{Quentin Remy\,\orcidlink{0000-0002-4801-3199}}
\affiliation{Department of Physics, Freie Universit{\"a}t Berlin, 14195 Berlin, Germany}

\author{Guillermo Nava Antonio\,\orcidlink{0000-0003-2813-2841}}
\affiliation{Microelectronics Group, Cavendish Laboratory, University of Cambridge, J.J.Thomson Avenue, Cambridge CB3 0HE, United Kingdom}

\author{Obed Alves Santos\,\orcidlink{0000-0002-0192-8236}}
\affiliation{Microelectronics Group, Cavendish Laboratory, University of Cambridge, J.J.Thomson Avenue, Cambridge CB3 0HE, United Kingdom}

\author{Michel Hehn\,\orcidlink{0000-0002-4240-5925}}
\affiliation{Universit{\'e} de Lorraine, CNRS, IJL, F-54000 Nancy, France}

\author{Richard F.L. Evans\,\orcidlink{0000-0002-2378-8203}}
\affiliation{School of Physics, Engineering and Technology, University of York, York, YO10 5DD, United Kingdom}

\author{Roy W. Chantrell\,\orcidlink{0000-0001-5410-5615}}
\affiliation{School of Physics, Engineering and Technology, University of York, York, YO10 5DD, United Kingdom}

\author{St{\'e}phane Mangin\,\orcidlink{0000-0001-6046-0437}}
\affiliation{Institut Jean Lamour (UMR 7198), Universit{\'e} de Lorraine, Vandoeuvre-l{\`e}s-Nancy, France}

\author{Christopher H. Marrows\,\orcidlink{0000-0003-4812-6393}}
\affiliation{School of Physics and Astronomy, University of Leeds, Leeds LS2 9JT, United Kingdom}

\author{Joseph Barker\,\orcidlink{0000-0003-4843-5516}}
\email{j.barker@leeds.ac.uk}
\affiliation{School of Physics and Astronomy, University of Leeds, Leeds LS2 9JT, United Kingdom}

\author{Chiara Ciccarelli\,\orcidlink{0000-0003-2299-3704}}
\email{cc538@cam.ac.uk}
\affiliation{Microelectronics Group, Cavendish Laboratory, University of Cambridge, J.J.Thomson Avenue, Cambridge CB3 0HE, United Kingdom}

\maketitle

\section*{S1. Temperature dependence of the correction function}

The temperature dependence of the THz emission in FeRh-Pt is the central part of our study. It is therefore very important to understand how the optical and electrical  properties of our sample depend on temperature to correctly account for changes in the absorbed pump fluence, $A(T)$, and the efficiency of THz field outcoupling $C(T)$. To filter out the contribution of these two variables from our temperature analysis, we introduce a correction function, $G(T)=1/(A(T)C(T))$, and define a renormalised THz emission amplitude $S^* _y(T)=G(T)S_y(T)$, where $S_y(T)$ is the measured amplitude. 

\begin{figure}[h]
    \centering
    \includegraphics[width=0.7\textwidth]{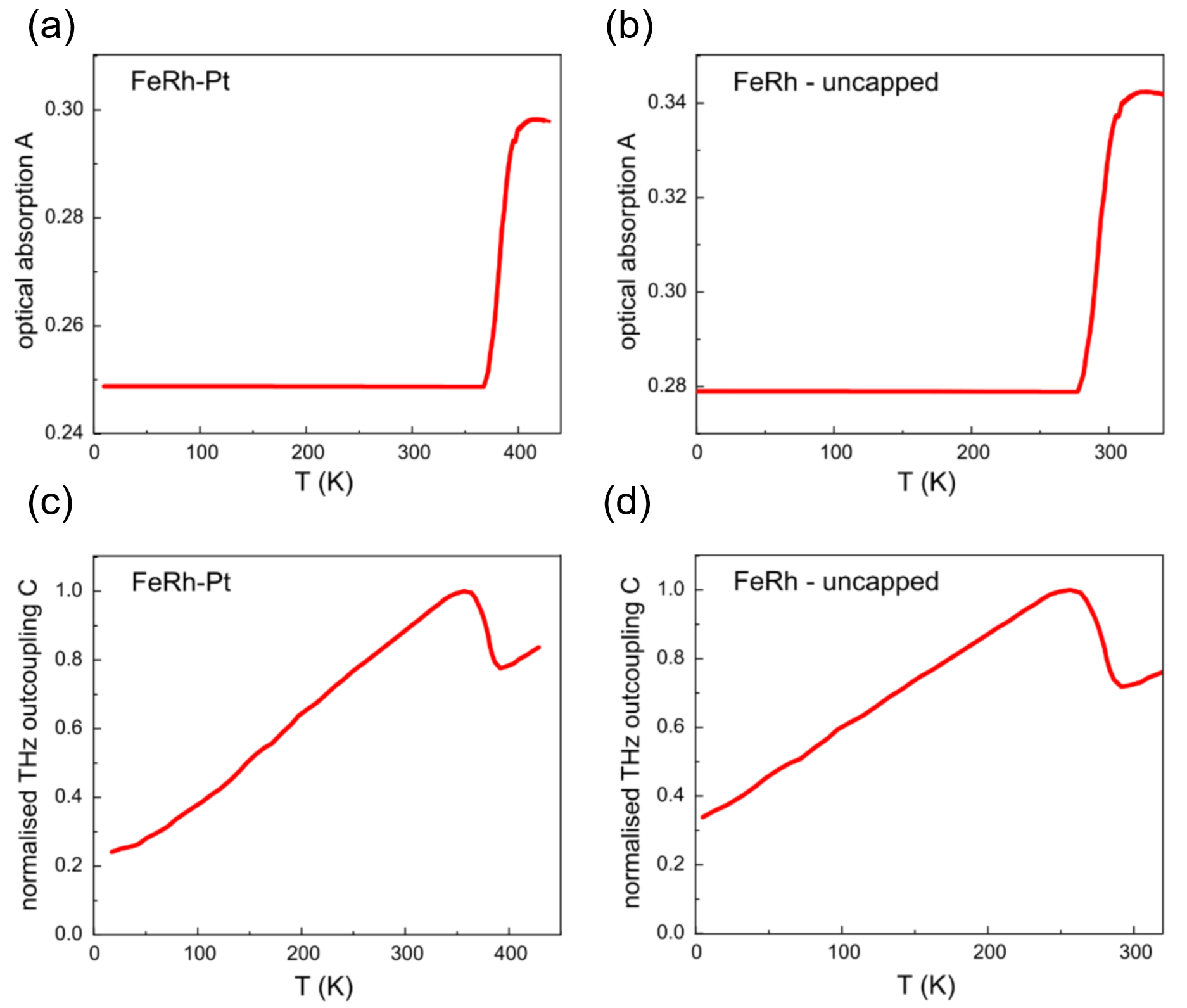}
    \caption
{
 Temperature dependence of the correction function parameters in the top pumping geometry: optical absorption fraction $A(T)$ \textbf{(a)} in FeRh(30)/Pt(3.5) and \textbf{(b)} uncapped FeRh (30); normalised THz outcoupling, $C(T)$ in \textbf{(c)} FeRh(30)/Pt(3.5) and \textbf{(d)} uncapped FeRh (30).
}
    \label{fig:figScorrection}
\end{figure}

To obtain $A(T)$ we calculate the total absorption of an optical laser light at 800 nm using the standard Transfer Matrix Method \cite{Byrnes2016}. To do so, we require the refractive indices of each of the constituents of the sample stack for the corresponding optical wavelength. We take the refractive indices of MgO and Pt to have negligible dependence on temperature as $n_\mathrm{MgO}($800 nm$)=1.7$ \cite{Stephens_JResNatlBureauStand_49_249_1952} and $n_\mathrm{Pt}($800 nm$)=0.95+4.71i$ \cite{Werner_JPhysChemRefData_38_1013_2009}. The temperature dependence of the refractive index of FeRh at the fixed angular frequency $\omega =2 \pi c/\lambda$ with $\lambda=800$ nm, is given by \cite{Saidl_NewJPhys_18_083017_2016}: 
\begin{equation}
\label{eqn:EqS1}
    n_{FeRh}^2(T)=1-\frac{\omega_p(T)^2}{\omega^2+1/\tau(T)^2}+\frac{i\omega_p(T)^2/\omega\tau(T)}{\omega^2+1/\tau(T)^2}+\frac{i\sigma_\mathrm{inter}(T)}{\varepsilon_0\omega}.
\end{equation}
$\tau(T)=\sigma_\mathrm{DC}(T)/(\varepsilon_0\omega_p^2(T))$ is the scattering time \cite{Saidl_NewJPhys_18_083017_2016}. The temperature dependence of the DC conductivity of FeRh, $\sigma_\mathrm{DC}(T)$, is obtained from van der Pauw measurements.  $\omega_p(T)$ is the plasma frequency, which we take as $\hbar\omega_p^\mathrm{AFM}=1.8$ eV and $\hbar\omega_p^\mathrm{FM}=5.5$ eV \cite{Saidl_NewJPhys_18_083017_2016} in the AF and FM phases of FeRh, respectively. $\sigma_\mathrm{inter}(T)$ is the interband conductivity, which we take as $\sigma_\mathrm{inter}^\mathrm{AFM}=-i\omega\varepsilon_0\times(0.74+0.58i)$ and $\sigma_\mathrm{inter}^\mathrm{FM}=-i\omega\varepsilon_0\times(18.26+32.77i)$ \cite{Vyborny}. 

To describe $\omega_p(T)$ and $\sigma_\mathrm{inter}(T)$ in the phase-transition region, we introduce a temperature dependent scaling parameter $\alpha(T)$, with $\alpha(T\ll T_\mathrm{(AF-FM)})=1$ in the AFM phase and $\alpha(T \gg T_\mathrm{(AF-FM)})=0$ in the FM phase. The temperature dependence of $\alpha$ is obtained from the renormalisation of the THz transmission data around the AFM/FM transition. The full temperature dependence of the plasma frequency and interband conductivity is then:
\begin{equation}
\label{eqn:EqS2}
    \omega_p(T)=\alpha(T)\omega_p^\mathrm{AFM}+(1-\alpha(T))\omega_p^\mathrm{FM},
\end{equation}
and
\begin{equation}
\label{eqn:EqS3}
    \sigma_{\mathrm{inter}}(T)=\alpha(T)\sigma_\mathrm{inter}^\mathrm{AFM}+(1-\alpha(T))\sigma_\mathrm{inter}^\mathrm{FM}. 
\end{equation}
The optical absorption temperature dependencies for the FeRh(30)/Pt(3.5) and uncapped FeRh(30) samples in the top-pumping geometry are plotted in Fig. S\ref{fig:figScorrection}(a) and (b) respectively.
These $A(T)$ find that for the maximum pump incident fluence of 9.52 mJ/cm$^2$, the absorbed fluences are in the ranges of 2.38-2.86 mJ/cm$^2$ and 2.67-3.28 mJ/cm$^2$ for FeRh(30)/Pt(3.5) and FeRh(30) respectively.

To find the temperature dependence of the THz field outcoupling at 1 THz, we use the expression given in Ref.~\onlinecite{Seifert_NatPhotonics_10_483_2016}: 
\begin{equation}
\label{eqn:EqS4}
   C(T)\propto\frac{1}{1+n_\mathrm{MgO}+Z_0\int_{0}^{d}{dz\ \sigma(T,z)}} .
\end{equation}
Here, $n_\mathrm{MgO}=3.1$ \cite{Nenno_SciRep_9_13348_2019} is the refractive index of the MgO substrate at 1 THz, considered to be temperature-independent. $Z_0=377$ $\Omega$ is the vacuum impedance, $d$ is the metal stack thickness, and $\sigma(T,z)$ is the conductivity distribution of the metal stack. At THz frequencies $\int_{0}^{d}{dz\ \sigma(T,z)}= d_\mathrm{FeRh}\sigma_\mathrm{FeRh}(T)+d_\mathrm{Pt}\sigma_\mathrm{Pt}(T)$, with $d_\mathrm{FeRh}= 30$ nm, and $d_\mathrm{Pt}= 3.5$ nm. The temperature dependence of the DC conductivities in FeRh \cite{Vries} and Pt \cite{Wang_ApplPhysLett_105_152412_2014} are obtained from the literature. The normalised $C(T)$ for the FeRh(30)/Pt(3.5) and FeRh(30) samples are plotted in Fig. S\ref{fig:figScorrection}(c) and (d) respectively. 

\section*{S2. Temperature dependence of the magnetisation}
\subsection{Magnetic hysteresis loops of FeRh(30)/Pt(3.5)}

\begin{figure}[h]
    \centering
    \includegraphics[width=1\textwidth]{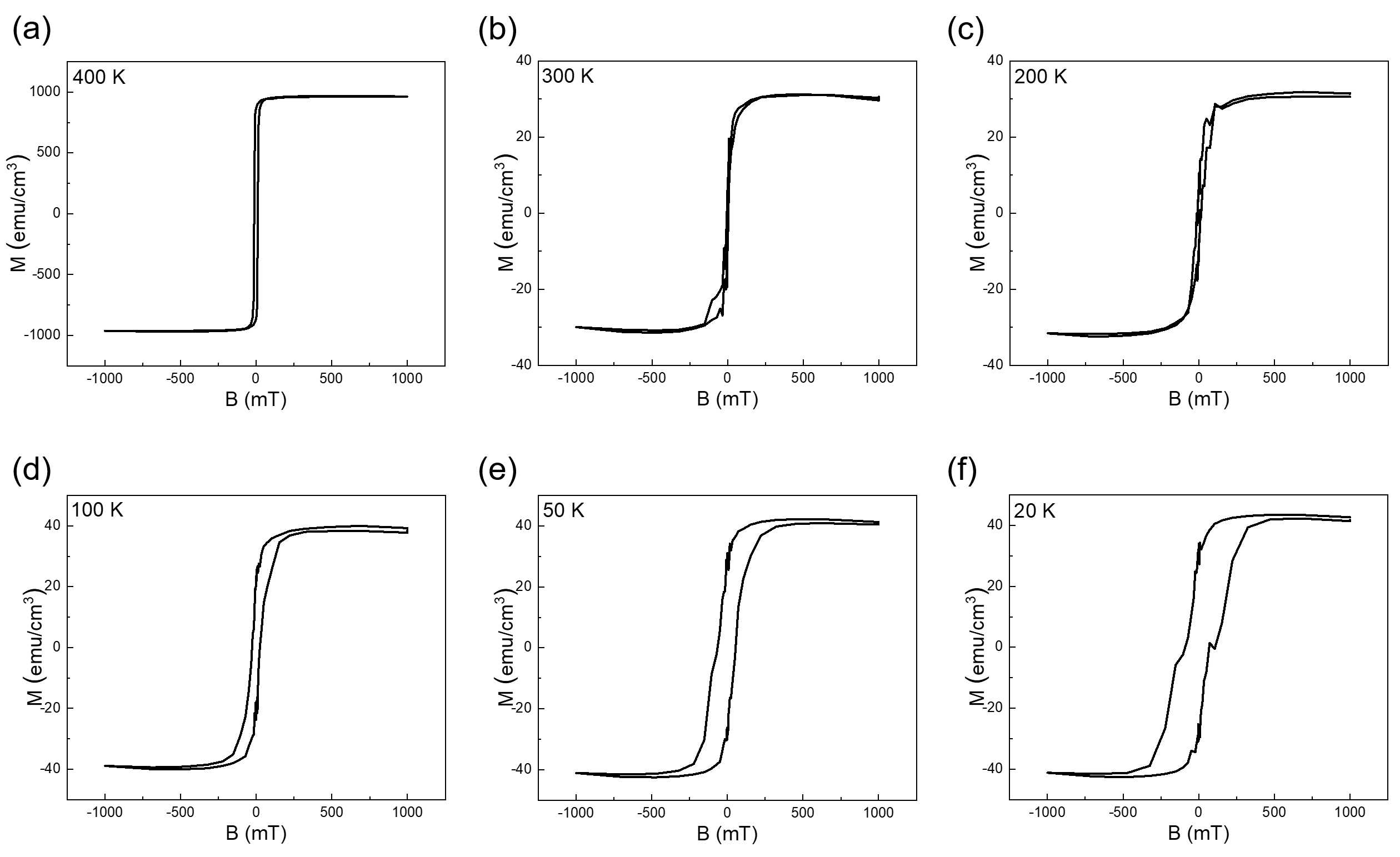}
    \caption
{
Magnetic hysteresis loops of FeRh(30)/Pt(3.5) measured by SQUID at 400 K \textbf{(a)}, 300 K \textbf{(b)}, 200 K \textbf{(c)}, 100 K \textbf{(d)}, 50 K \textbf{(e)} and 20 K \textbf{(f)}, after subtracting the paramagnetic contribution and the contribution of the MgO(001) substrate. The saturation magnetisation and coercive field values in the main article were acquired from this data set.
}
    \label{fig:figS1}
\end{figure}

Fig. S\ref{fig:figS1} shows hysteresis loops of FeRh(30)/Pt(3.5) measured using a SQUID magnetometer at temperatures between 400 K and 20 K. The MgO(001) background was removed using data sets obtained for a bare substrate. The paramagnetic contributions were subtracted. From these measurements we extract the temperature dependence of the saturation magnetisation $M_\mathrm{sat}$, shown in Fig. 1(b) and 2(d) of the main text and of the coercive $B_c$ and exchange bias $B_\mathrm{ex}$ fields in Fig. 2(d) of the main text. 

\subsection{Four-spin atomistic spin model}
\label{sec:four-spin-model}

To model FeRh we used the atomistic spin dynamics software package \vampire \cite{Evans_JPhysCondensMatter_26_103202_2014, *vampire} that describes the dynamics of each spin, $\mathbf{S}_i$. The metamagnetic transition of FeRh is modelled in terms of the competition between the higher-order exchange interaction (four-spin) responsible for the antiferromagnetic ordering (AFM) at low temperatures and the bilinear exchange interaction, responsible for ferromagnetic ordering (FM) at elevated temperatures. The model has been developed by Barker and Chantrell \cite{barker} and further used in the literature \cite{ostler2017modeling, strungaru_physrevmat}.

The spin Hamiltonian used in the atomistic simulations includes contributions from the bilinear exchange in form of nearest and next-nearest neighbors exchange and four-spin interaction and has the following form:
\begin{equation}
\begin{split}
\label{eq::gen_ham}
{\mathcal{H}} =&-\frac{1}{2}\,  \sum_{i,j } J_{ij}\, (\mathbf{S}_i\, \cdot  \mathbf{S}_j\:) -k_u \sum_{i} (\mathbf{S}_i\, \cdot  \mathbf{e})^2 \\
&-\frac{1}{4}\,  \sum_{i,j,k,l } \frac{1}{3}D_{ijkl}\, \left( (\mathbf{S}_i\, \cdot  \mathbf{S}_j\:)(\mathbf{S}_k\, \cdot  \mathbf{S}_l\:) + (\mathbf{S}_i\, \cdot  \mathbf{S}_k\:)(\mathbf{S}_j\, \cdot  \mathbf{S}_l\:)+ (\mathbf{S}_i\, \cdot  \mathbf{S}_l\:)(\mathbf{S}_k\, \cdot  \mathbf{S}_j\:)\right),
\end{split}
\end{equation}
where $J_{ij}$ and $D_{ijkl}$ represent  the bilinear and four-spin exchange interaction between Fe atomic sites, respectively, $k_\mathrm{u}$ represents the uniaxial anisotropy constant, with $\mathbf{e}$ representing the easy axis direction. The factor $\frac{1}{2}$ accounts for the double summation $i \rightarrow j$ and $j \rightarrow i$ since numerically both interactions are included explicitly. Similarly, for the four-spin exchange term, the factor $\frac{1}{4}$ appears due to the explicit inclusion of all interactions for atoms $i$, $j$, $k$, $l$. 

%



%

The parameters used in the simulations are presented in Table \ref{parameters_simulation}, with the exchange parameters being summarised in Table \ref{tab:fourspin_exchange}. The value of the uniaxial anisotropy was extracted from Ref.~\cite{ostler2017modeling}. We calculate the equilibrium magnetisation of the system by employing Monte-Carlo simulations, where we equilibrate the system at each temperature for 50,000 Monte-Carlo steps, and then average the properties for 50,000 more steps. 

\begin{figure}[h]
    \centering
    \includegraphics[trim={0 0cm 0cm 0},width=0.7\textwidth]{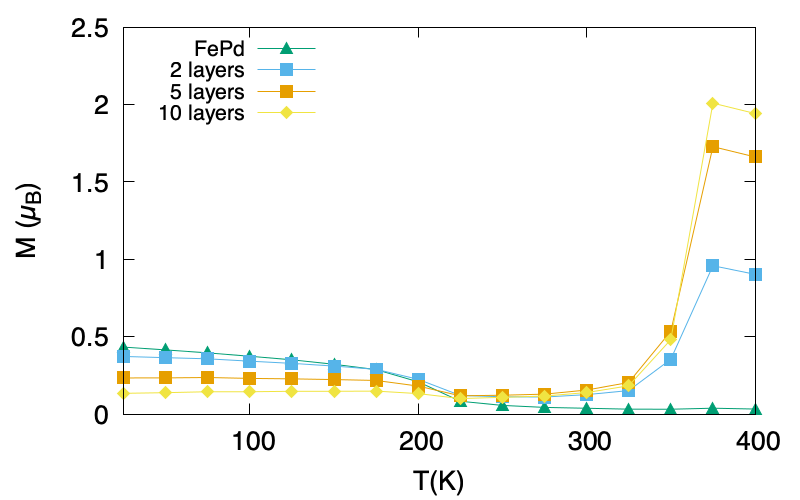}
    \caption
{ Temperature dependence of magnetisation as extracted from atomistic spin dynamics simulations. The green curve shows the magnetisation of only the FePd monolayer. The blue, orange and yellow curves show the temperature dependence of the total magnetisation of 2, 5 and 10 monolayes of FeRh respectively plus a monolayer of FePd.
}
    \label{fig:figS1_th}
\end{figure}

We modelled a system of 65 $\times$ 32 $\times $ 32 atomic sites, periodic in the $yz$ directions, where at the FeRh/Pt interface we assume a FM monolayer of FePd. A reduced exchange ($J_\mathrm{FePd}$) compared to the bulk FePd is used for the monolayer, an effect occuring due to the disordered state of the interface. The magnetic moment at the interface is also reduced to $0.45~\mu_B$, a reduction in the moment being also observed in literature \cite{fan_ferh}. From the SQUID data $M_\mathrm{sat}$ (20 K) is about 20 times smaller than $M_\mathrm{sat}$ (400 K), renormalising by the total volume of FeRh. If we consider that the FM interfaces extend over a combined thickness of 10 nm, we obtain that the magnetisation of the interfaces at low temperature is about 7 times smaller than in the ferromagnetic phase, hence the value of $0.45~\mu_B$. This is just a rough approximation because the ferromagnetic interface with the MgO substrate is of a different nature and its magnetic moment might be different. A coupling of $10\%$ of the exchange value of FePd is used across the interface. The inclusion of the FM monolayer leads to a similar temperature dependence of magnetisation as observed experimentally, with a low Curie temperature of the monolayer (of 220K)  - green curve in Fig. S\ref{fig:figS1_th}.

\begin{center}
\begin{table}[ht!]
\centering
\caption{Parameters used for the simulation of FeRh/Pt.}
\begin{tabular}{c|lll}
\hline \hline
Quantity  & Symbol  & Value   &   Units   
 \\ \hline 

FeRh anisotropy energy  & $K_\mathrm{FeRh}$  & $1.404 \times 10^{-23}$    & J        \\
FePd anisotropy energy  & $K_\mathrm{FePd}$  & $2.63 \times 10^{-22}$    & J        \\
FeRh magnetic moment & $\mu_\mathrm{FeRh}$  & $3.15$    &  $\mu_\mathrm{B}$       \\
FePd magnetic moment & $\mu_\mathrm{FePd}$  & $0.45$    &  $\mu_\mathrm{B}$       \\
 \hline
\hline
\end{tabular}

\label{parameters_simulation}
\end{table}
\end{center}

\begin{table*}[h]
\caption{\label{tab:fourspin_exchange} Exchange constants for the four-spin exchange Hamiltonian~\eqref{eq::gen_ham}.}
\begin{ruledtabular}
\begin{tabular}{lcccdd}
    symbol   & types  & number & \text{distance (lattice constants)} & \text{value (meV)} \\ \hline
    $J_{2}^\mathrm{FeRh}$    & Fe-Fe &$6$ &$1$ & $2.49$\footnotemark[1]  \\
    $J_{3}^\mathrm{FeRh}$    & Fe-Fe & $12$  & $\sqrt{2}$ & $17.16$\footnotemark[1]\\
    $J_{2}^\mathrm{FePd}$    & Fe-Fe & $6$  & $1$ & $2.49$  \\
    $J_{2}^\mathrm{FeRh-FePd}$    & Fe-Fe & $6$  & $1$ & $0.249$\\
    \colrule
    $D_{ijkl}^\mathrm{FeRh}$    & Fe-Fe-Fe-Fe & 32&1& $1.43$\footnotemark[1]
\end{tabular}
\end{ruledtabular}
\footnotetext[1]{Ref. \onlinecite{barker}}
\end{table*}

We next investigate the magnetisation behaviour arising from the various thickness of the considered interface. For an interface consisting of 2 layers (monolayer of FePd and monolayer of FeRh) blue curve in Fig. \ref{fig:figS1_th}) we observe that the magnetisation in the low temperature regime has a similar amplitude that the one arising after the first order phase transition in FeRh. By distancing from the FePd interface, a decrease in the low temperature magnetisation is observed (as shown for 5 and 10 layers, orange and yellow curve).   

The field dependence observed experimentally can appear as an effect of magnetic domains. For the simulation results presented above, the system size considered allows only for monodomain configurations at low temperature. However, for larger system sizes (which go beyond our computational capabilities), magnetic domains can be present at low temperature. Applying a magnetic field will saturate these domains in the direction of the field, hence the increased in the overall magnetisation along field direction can lead to an increse in the spin current with enhanced field as shown experimentally. 

\section*{S3. Additional experimental data}

\subsection{Raw THz emission data of FeRh(30)/Pt(3.5)}

\begin{figure}[h]
    \centering
    \includegraphics[width=1\textwidth]{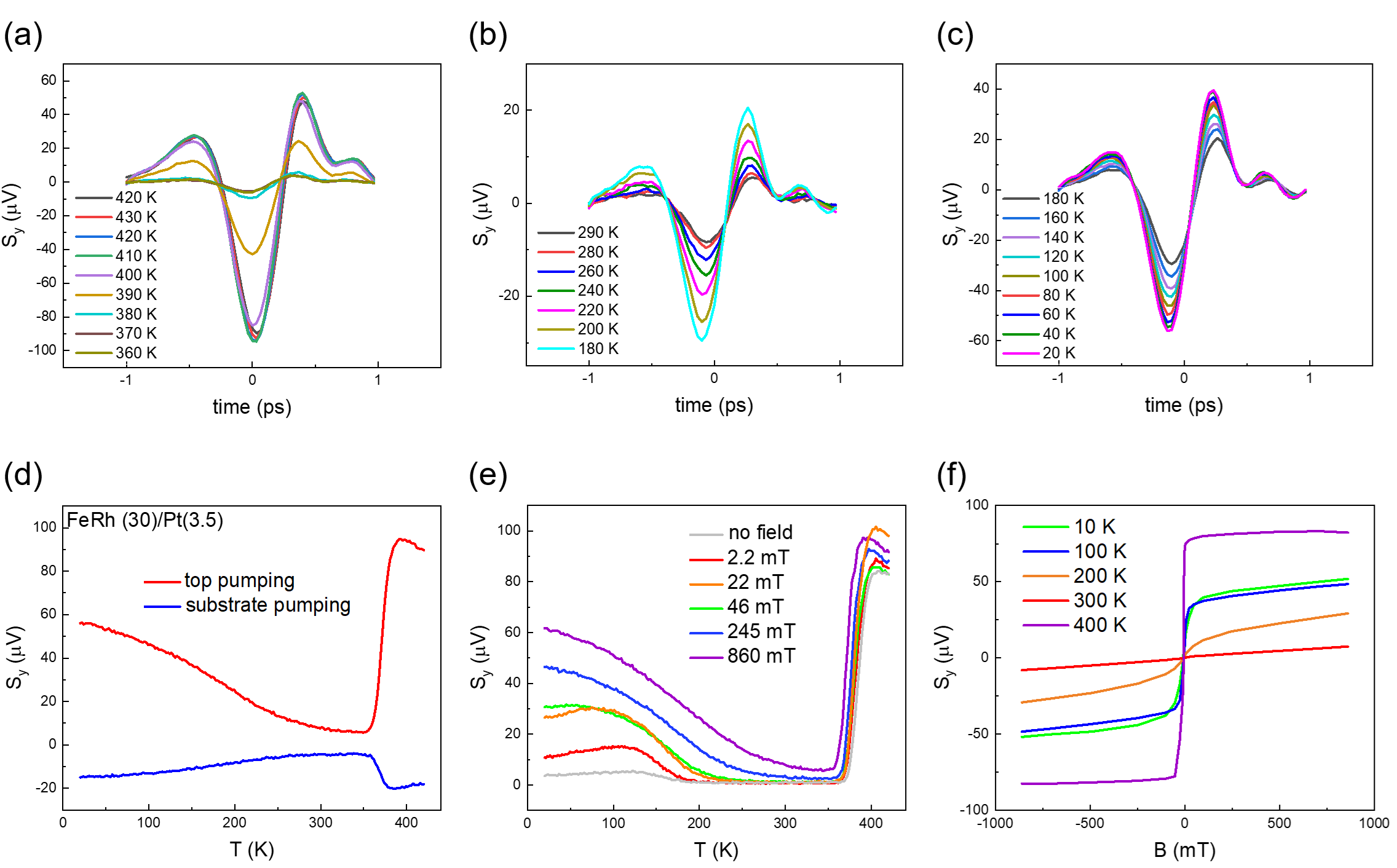}
    \caption
{\textbf{(a)-(c)} THz transient time-traces measured at different temperatures in the range 420 K-360 K \textbf{(a)}, 290 K-180 K \textbf{(b)} and 180 K - 20 K \textbf{(c)}. \textbf{(d)} Temperature dependence of the THz emission amplitude measured at 860 mT for two pumping geometries, pumping the sample from the Pt side and measuring the THz emission behind the substrate (red), or pumping the sample from the substrate side and measuring the THz emission behind the Pt layer (blue).
\textbf{(e)} Temperature dependence of the THz emission amplitude for different values of the in-plane magnetic field. \textbf{(f)} Dependence of the THz emission amplitude on the in-plane magnetic field at different temperatures. For these measurements a top pumping geometry was adopted, with the pump hitting the sample from the Pt side and the THz emission being measured behind the substrate.
}
    \label{fig:figS5}
\end{figure}

Fig. S\ref{fig:figS5} presents the temperature and magnetic field dependence of the raw THz emission data, before renormalising by the temperature dependent optical absorption and temperature dependent THz outcoupling. The THz emission amplitude undergoes an abrupt decrease across the phase transition from ferromagnetic to antiferromagnetic. As temperature is further decreased, the THz emission amplitude starts increasing again up to 60\% of the value in the ferromagnetic phase. 

\subsection{Magnetic properties and raw THz signal of uncapped FeRh(30 nm)}

\begin{figure}[h]
    \centering
    \includegraphics[width=0.7\textwidth]{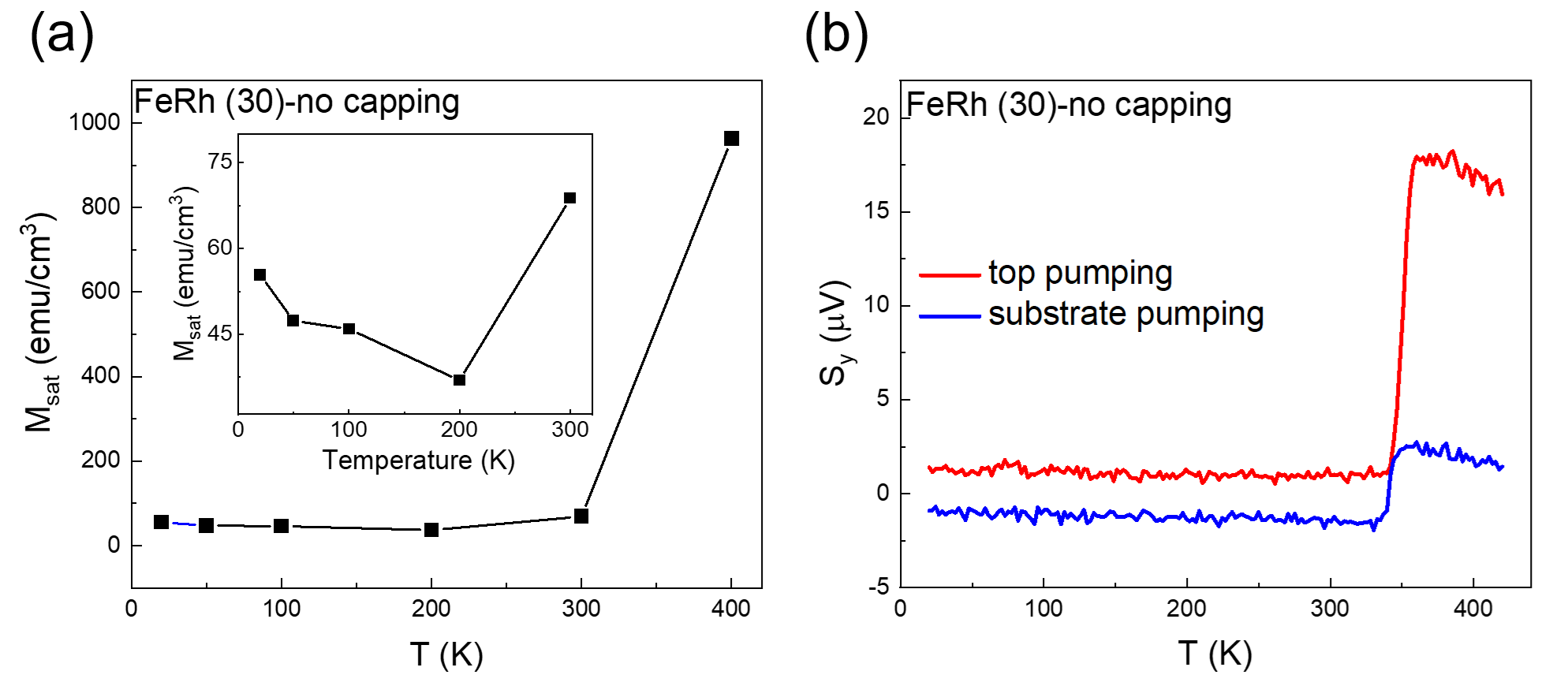}
    \caption
{
\textbf{(a)} Temperature dependence of the saturation magnetisation $M_{sat}$ extracted from SQUID magnetometry measurements for an uncapped FeRh(30 nm) film. The inset represents a zoom-in of the low temperature region below 300 K. \textbf{(b)} Temperature dependence of the THz emission amplitude in uncapped FeRh(30 nm). An optical fluence of 1.2 mJ/cm$^2$ was used and an in-plane field of 860 mT was applied during the measurement. The two curves represent measurements performed for the sample pumped from the substrate side (blue curve) and top surface (red curve).
}
    \label{fig:figS6}
\end{figure}

The SQUID data in Fig. S\ref{fig:figS6} (a) shows that in the uncapped FeRh(35) sample, residual ferromagnetism exists also in the antiferromagnetic phase and its saturation magnetisation reaches values comparable to the FeRh(30 nm)/Pt(3.5 nm) sample at low temperature. Similarly to the FeRh/Pt sample, we see that the residual magnetisation is never zero but starts increasing slightly below 200 K. Polarized neutron reflectivity measurements have shown that this residual magnetism mainly resides at the interface with the MgO substrate and the top interface. Residual magnetism at the bottom interface does not have a strong dependence on temperature \cite{Almeida_PhysRevMater_4_034410_2020}. Residual magnetism at the top surface, which is of interest for the interpretation of our THz emission measurements, can be caused by alloying and therefore depends on the capping material and unsually has a stronger dependence on temperature \cite{fan_ferh, Baldasseroni_ApplPhysLett_100_262401_2012, Ding_JApplPhys_103_07B515_2008}. In our case, however, we don not observe a significant difference in the SQUID data of capped and uncapped FeRh and we therefore tend to associate the ferromagnetic top-layer with ferromagnetic FePd that forms independently on capping. Previous works have shown that for uncapped FeRh and FeRh capped with Pt, ferromagnetism at the top-surface is negligible  \cite{Baldasseroni2014, Baldasseroni_ApplPhysLett_100_262401_2012}. Differently from these works, our FeRh is doped with 2.8\% Pd at the top interface.  

Fig. S\ref{fig:figS6} (b) shows the THz emission amplitude as a function of temperature. As discussed in the main text, in the FM phase the THz emission has a magneto-dipole nature due to the reduced contribution from the spin-Hall effect. In the AF phase we still observe a non-zero THz emission, odd with respect to sample flipping. We explain this with the non-negligible spin-Hall angle of AF-FeRh.

\subsection{Optical fluence dependence of THz emission from FeRh(30)/Pt(3.5)}

\begin{figure}[H]
    \centering
    \includegraphics[width=0.7\textwidth]{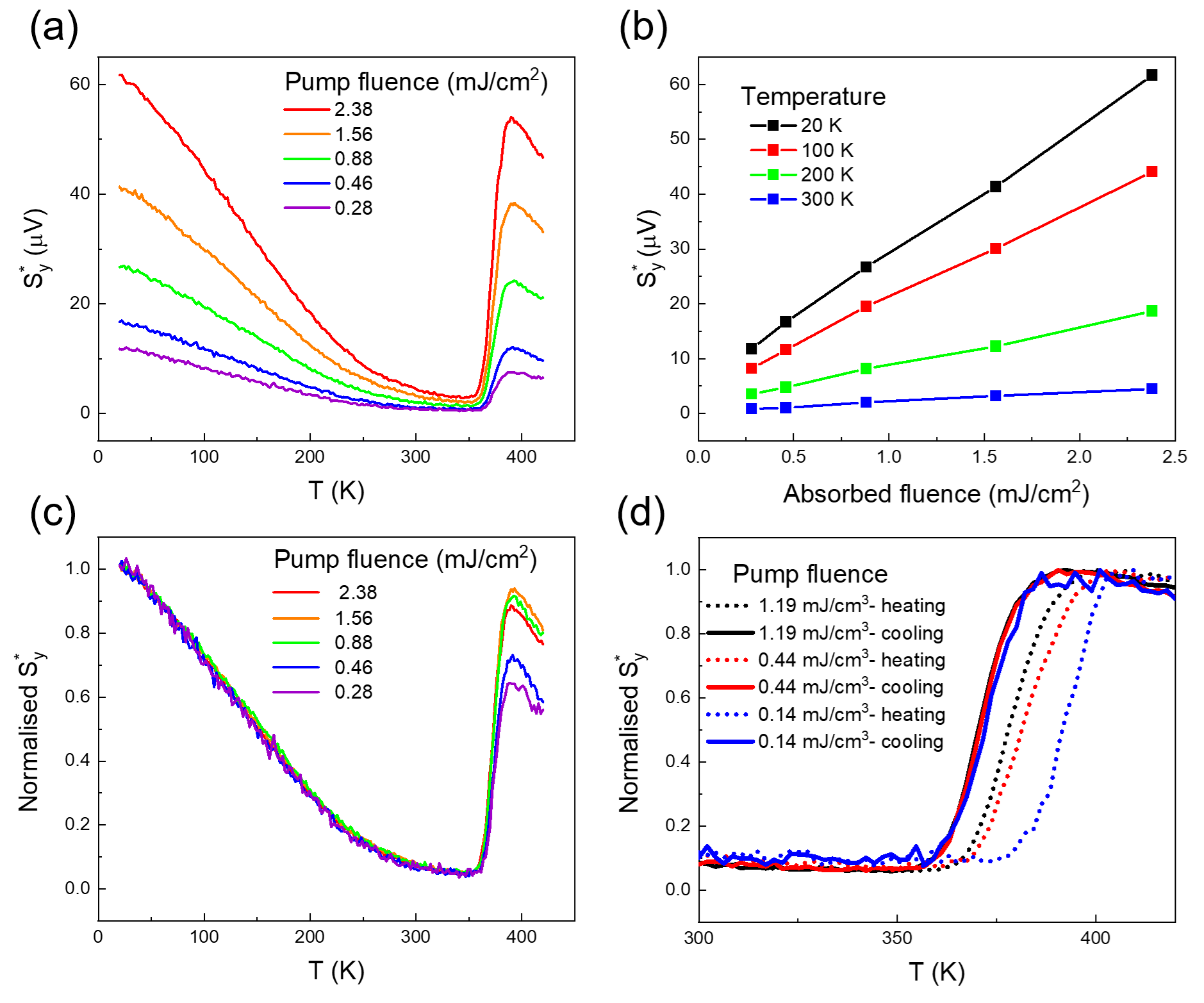}
    \caption
{
Optical fluence dependence of THz emission from FeRh(30)/Pt(3.5). \textbf{(a)} Temperature dependence of $S_y^*$ for different pump fluences between $0.28$ and $2.38$ mJ/cm$^2$. \textbf{(b)} $S_y^*$ plotted as a function of pump fluence at selected temperatures between 20 and 300 K. \textbf{(c)} Data sets from \textbf{(a)} normalised with respect to the values of $S_y^*$ at 20 K. \textbf{(d)} Temperature dependence of normalised $S_y^*$  for different pump fluences. Both cooling and heating branches are shown.  
}
    \label{fig:figfluence}
\end{figure}

Fig. S\ref{fig:figfluence} shows how the THz emission from FeRh(30)/Pt(3.5) depends on the optical pump fluence. Normalised temperature dependence data presented in (c) demonstrates that the low temperature THz emission enhancement is qualitatively identical for all studied fluences as all curves overlap in the 20-300 K range. In (b), we show that the magnitude of $S_y^*$ is linear with the fluence at temperatures in the 20-300 K range, and that no threshold behaviour is observed. These results exclude that the THz enhancement at low temperatures is due to a phase transition induced by transient pump heating. In Fig. S6(d) we focus instead on the FM-AF transition region and show that as the pump fluenece is increased the hysteresis closes, in agreement with previous works \cite{Awari_ApplPhysLett_117_122407_2020}.  

\subsection{Dependence of THz emission on FeRh doping in FeRh(30)/Pt(3.5) structures}

According to our microscopic model, residual ferromagnetism at the interface with Pt plays a key role in triggering spin pumping from antiferromagnetic FeRh. This ferromagnetism will depend on the chemical composition and doping level of the interface \cite{PdDoping}. We have therefore compared spin pumping in two different samples with the following composition: 

  \item \textbf{FeRh35-Pt}: \text{MgO/Fe}_{50}\text{Rh}_{46.8}\text{Pd}_{1.7}\text{Ir}_{1.5}(5)/\text{Fe}_{50}\text{Rh}_{47.1}\text{Pd}_{2.2}\text{Ir}_{0.7}(10)/\text{Fe}_{50}\text{Rh}_{47.2}\text{Pd}_{2.8}(15)/\text{Pt}(3.5)}
  
  \item \textbf{FeRh36-Pt}:   \text{MgO/Fe}_{50}\text{Rh}_{47.2}\text{Pd}_{2.8}(15)/ \text{Fe}_{50}\text{Rh}_{47.1}\text{Pd}_{2.2}\text{Ir}_{0.7}(10)/ \text{Fe}_{50}\text{Rh}_{46.8}\text{Pd}_{1.7}\text{Ir}_{1.5}(5)/\text{Pt}(3.5)}
\\

\begin{figure}[H]
    \centering
    \includegraphics[width=0.7\textwidth]{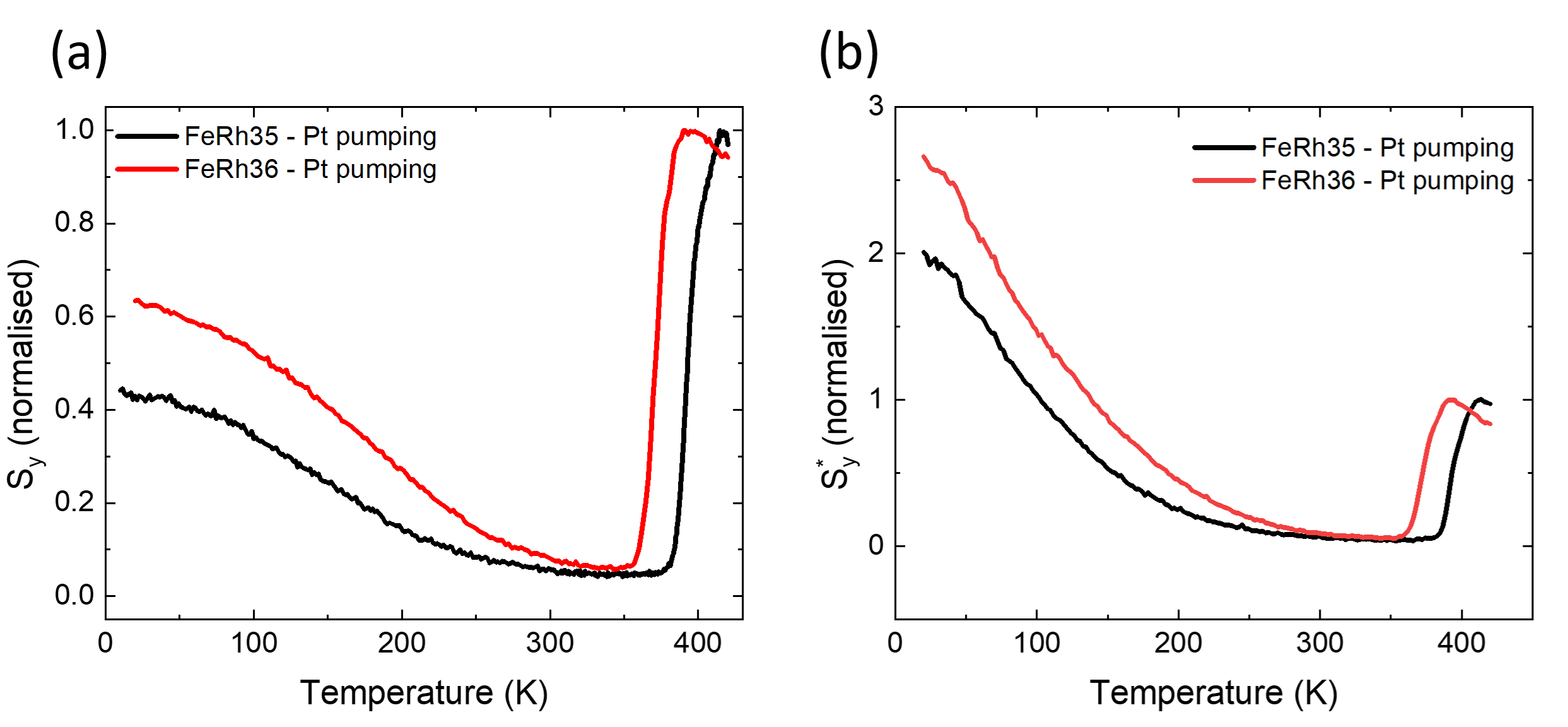}
    \caption
{Temperature dependence of $S_y$ \textbf{(a)} and $S_y^*$ \textbf{(b)} when pumping from the Pt side in sample FeRh35-Pt (black) and FeRh36-Pt (red). The pump fluence is 2.38 mJ/cm$^2$ and a magnetic field of 860 mT is applied.
}
\label{doping}
\end{figure}

The two samples have identical doping levels but the doping gradient is inverted. In this way we guarantee very similar bulk properties (crystal quality and conductivity, which determines the THz outcoupling), but different interface properties. In Fig. S\ref{doping} we compare the temperature dependence of the THz emission amplitude $S_y$ and spin pumping $S^*_y$ in the two samples, normalised to the values in the ferromagnetic phase. First, we notice that in sample FeRh35-Pt, the phase transition is shifted to higher temperatures. This is in agreement with the fact that Ir-rich FeRh has a higher transition temperature than Pd-rich FeRh \cite{10.1063/1.4907282} and confirms that only the volume closer to the Pt interface is involved in the spin-pumping. Also, we see that as temperature is decreased below the transition temperature, we measure a lower value of the spin pumping in FeRh35-Pt. This shows that the chemical composition of the interface with Pt is important. The fact that spin pumping decreases with Pd concentration provides further evidence that residual magnetism resides in FePd alloying.

\subsection{THz transmission through FeRh(30)/Pt(3.5}

\begin{figure}[H]
    \centering
    \includegraphics[width=0.6\textwidth]{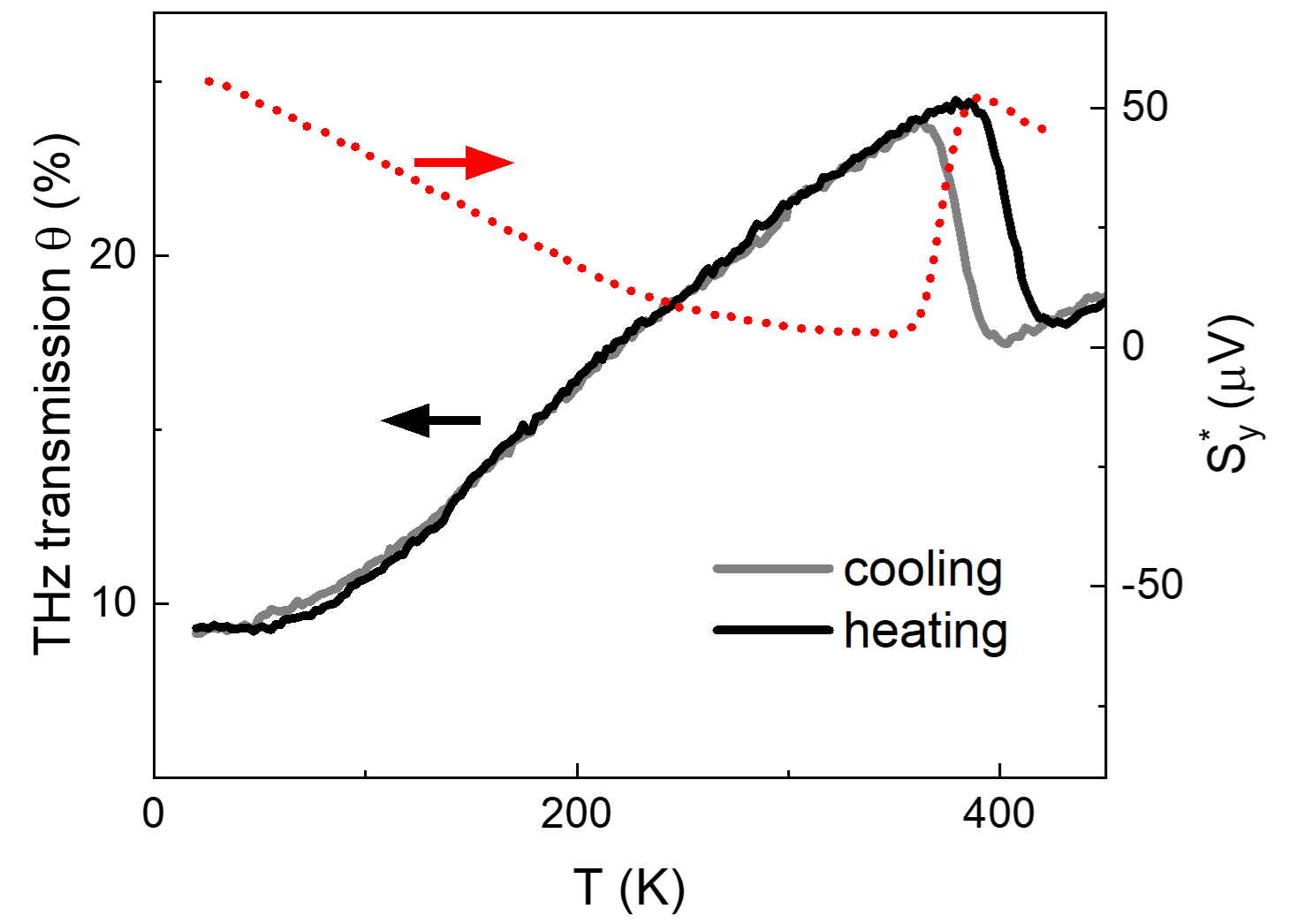}
    \caption
{Transmission of 1 THz field through FeRh(30)/Pt(3.5) for Pt incidence. Cooling (grey) and heating
(black) branches are indicated. The experiment was performed with an external magnetic field of 860 mT continuously
applied in-plane. The red dotted line indicates the temperature dependence of $S^*_y$ when cooling from high to low temperature. The pump fluence
is 2.38 mJ/cm2 and a magnetic field of 860 mT is applied.}
    \label{fig:figtrans}
\end{figure}

Fig. S\ref{fig:figtrans} shows the THz transmissivity of the FeRh-Pt sample as a function of temperature. $\theta$ is defined as the ratio between the transmitted THz radiation and the reference signal measured with no sample in the propagation path at 1 THz. Changes in the structural and electronic properties of FeRh across the AF-FM transition result in a steep variation of $\theta$ around $T_\mathrm{(AF-FM)}$. The phase transition region displays a clear temperature hysteresis between the cooling and heating branches, as expected for a first-order phase transition. Below $T_\mathrm{(AF-FM)}$, $\theta$ continuously decreases, which is attributed due to the increasing conductivities of FeRh and Pt \cite{Glover_PhysRev_108_243_1957, Huisman_PhysRevB_92_104419_2015}. 

\par

However, if we compare the value of $T_\mathrm{(AF-FM)}$ measured via THz transmission and optical pump-THz emission (red dotted line) we observe that this is about 10 K lower in the second case, which could be explained by a small amount of accumulated heat when the sample is pumped optically. 

\subsection{Comparison with MgO/CoFeB(30)/Pt(5) emitter}

\begin{figure}[h]
    \centering
\includegraphics[width=0.8\textwidth]{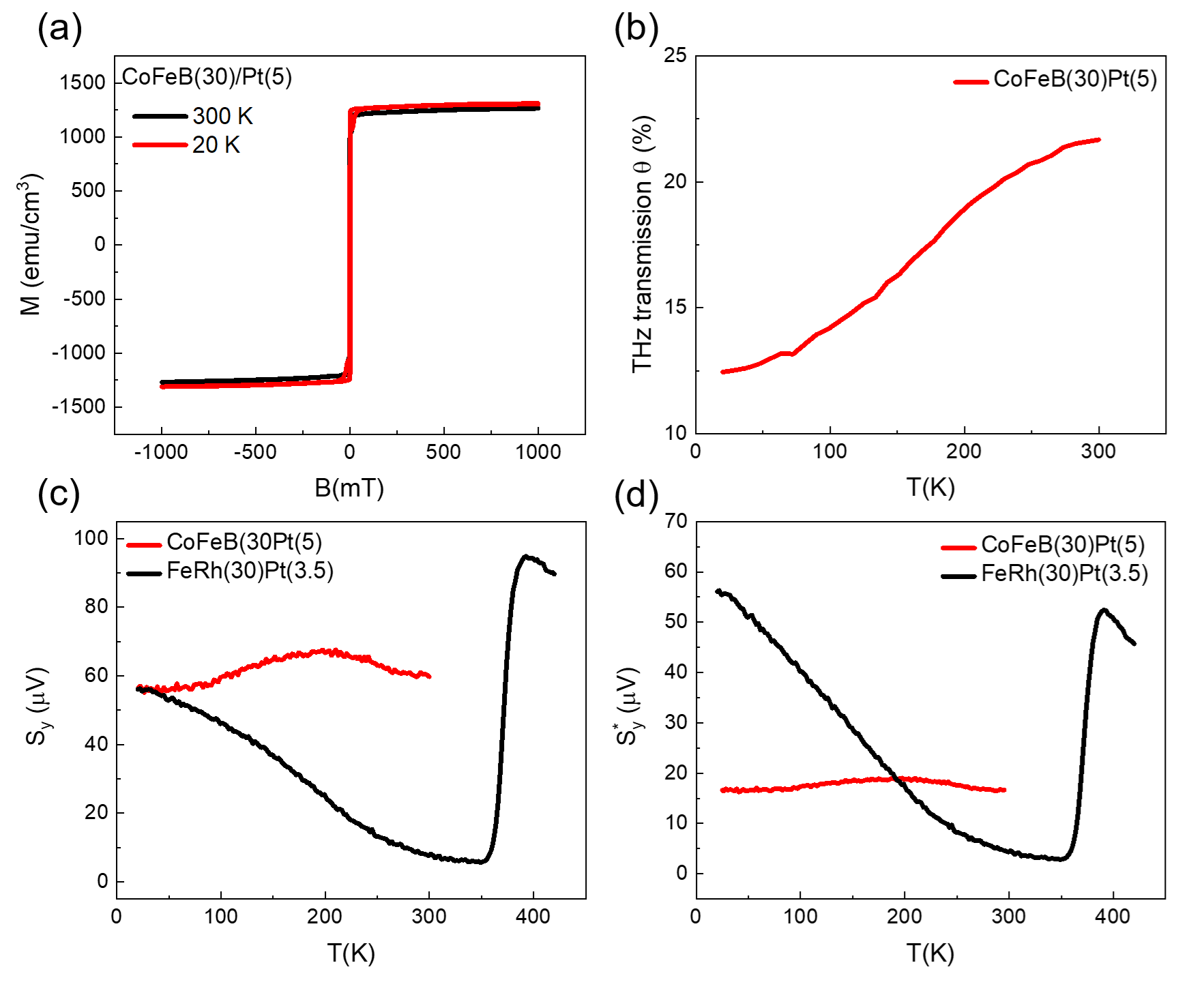}
    \caption{\textbf{(a)} Magnetic hysteresis loops of CoFeB(30)/Pt(5) measured by SQUID at 300 K (black) and 20 K (red).  \textbf{(b)} Transmission of 1 THz field through CoFeB(30)/Pt(5) for Pt incidence. The experiment was performed with an external magnetic field of 860 mT continuously
applied in-plane. \textbf{(c)} Temperature dependence of the THz emission amplitude $S^_{y}$ in CoFeB(30)/Pt(5) (red) and FeRh(30/Pt(3.5) (black). The measurement was done applying a 860
mT magnetic field and pumping the sample from the Pt side and measuring the THz emission behind the
substrate. \textbf{(d)} Temperature dependence of the THz emission amplitude rescaled by the optical absorption and THz outcoupling, $S^{*}_{y}$, in top-pumping geometry with an applied magnetic field of 860 mT for CoFeB(30)/Pt(5) (red) and FeRh(30)/Pt(3.5) (black).
}
       \label{fig:figS7}
\end{figure}

Here we compare the temperature dependence of the THz emission in FeRh(30)/Pt(3.5) and CoFeB(30)/Pt(5). The magnetisation of CoFeB varies only slightly with temperature in the explored temperature range (Fig. S\ref{fig:figS7} (a)), while the decrease in THz transmissivity as temperature is lowered reflects the increase in conductivity, similarly to the FeRh sample (Fig. S\ref{fig:figtrans}). The THz emission (Fig. S \ref{fig:figS7} (c)-(d)), although comparable in magnitude, follows a very different temperature dependence in the two samples, decreasing with decreasing temperature in the CoFeB sample. 

\section*{S4. Generalised Atomistic Spin Dynamics Model}

In the four-spin FeRh model in S2~\ref{sec:four-spin-model} the Rh moments are not explicitly represented, only the Fe moments are simulated. This means we cannot simulate the effect of Rh moments being induced by the laser pulse. We therefore develop a second spin model within the \textsc{JAMS} software package, where the dynamics of both the Fe and Rh moments are solved.

The FeRh crystal has the space group Pm$\overline{3}$m (221). We perform calculations using the cubic cell with the basis vectors
\begin{align}
\begin{split}
    \mathbf{a} &= a_0(1, 0 ,0), \\
    \mathbf{b} &= a_0(0, 1, 0), \\
    \mathbf{c} &= a_0(0, 0, 1),
\end{split}
\end{align}
with the lattice constant $a_0=3.0$~\AA. In the cubic cell, there is one Fe atom and one Rh atom. These are located at (in fractional coordinates)
\begin{align}
\begin{split}
    \text{Fe} &= (0.0, 0.0, 0.0), \\
    \text{Rh} &= (0.5, 0.5, 0.5).
\end{split}
\end{align}
The Hamiltonian is
\begin{equation}
    \mathcal{H} = -\frac{1}{2}\sum_{ij} J_{ij} \mathbf{S}_i \cdot \mathbf{S}_j - \frac{1}{2} \sum_{ij \in Fe} B_{ij}(\mathbf{S}_i \cdot \mathbf{S}_j)^2 -\sum_{i\in \mathrm{Fe}}k(\hat{\mathbf{e}}\cdot\mathbf{S}_i)^2 + \sum_{i} L_{2,i}  |S_i|^2 + L_{4,i}  |S_i|^4 + L_{6,i}  |S_i|^6,
    \label{eq:gse_hamiltonian}
\end{equation}
where $\vec{S}_i$ are dimensionless classical vectors which can change in length. The first term is the Heisenberg exchange, where $J_{ij}$ are the interaction energies. We used values from Polesya et al.~\cite{Polesya_PhysRevB_93_024423_2016} for Fe-Fe interactions and Fe-Rh interactions. Rh-Rh interactions are considered to be negligible. 
The second term in the Hamiltonian is a biquadratic exchange between Fe sites, which is necessary for the metamagnetic phase transition~\cite{Polesya_PhysRevB_93_024423_2016, Derlet_PhysRevB_85_174431_2012}. The factors of $1/2$ account for the double counting of each bond. The third term is the uniaxial anisotropy energy of the Fe sites with an axis $\hat{\mathbf{e}}$. The final term is a Landau Hamiltonian that describes the energetics of changes in the length of spin moments, where $L_2, L_4, L_6$ are parameters with units of energy. In our model of FeRh the Fe moments have a stable minimum at unit length. The Rh moments are unstable and contain only an $L_2$ term. This means that the minimum energy for the Rh Landau term is for zero spin length. However, the ferromagnetic Fe-Rh interactions compete with this and cause a Rh moment to form if there is a net Fe magnetisation. The values we used in our model Hamiltonian are given in Table~\ref{tab:asd_exchange}.

We use the generalised Langevin spin equation~\cite{Ma_PhysRevB_86_054416_2012}
\begin{equation}
    \frac{\partial \mathbf{S}_i}{\partial t} = - \gamma\mathbf{S}_i \times \mathbf{B}_i + \gamma\lambda\mathbf{B}_i + \bm{\xi}_i
    \label{eq:gse}
\end{equation}
where $\mathbf{B}_i = -(1/\mu_i)\nabla_{\mathbf{S}_i} \mathcal{H}$ is the effective field in teslas with $\mu_i$ the magnetic moment on site $i$ in Bohr magnetons, $\mu_{\mathrm{B}}$. $\mathbf{\xi}_i$ describes stochastic processes of the Langevin thermostat. $\lambda$ is a dimensionless damping parameter which describes the coupling to the bath. Here we use a quantum thermostat, the correlations of $\xi_i$ obey

\begin{equation}
    \langle \xi_{i,a}(t) \rangle = 0; \quad \langle \xi_{i,a}(t) \xi_{j,b}(t')\rangle_\omega = \delta_{ij}\delta_{ab}\frac{2\lambda k_{\mathrm{B}}T}{\gamma\mu_i}\frac{\hbar\omega}{\exp{(\hbar\omega/k_{\mathrm{B}} T)-1}},
\end{equation}
where $a,b$ are cartesian components ($x,y,z$), $k_{\mathrm{B}}$ is Boltzmann's constant, $T$ is the temperature in kelvins. $\langle \cdots \rangle_{\omega}$ indicates the correlation function is defined in frequency ($\omega$) space. Further details of the quantum thermostat implementation can be found in Ref.~\onlinecite{barker}. We integrate~\eqref{eq:gse} numerically using the Runge-Kutta fourth order method with a timestep of $\Delta t = 0.1$~fs. The values of the material parameters used for FeRh are given in Table~\ref{tab:asd_model_params}.

To simulate the effect of a transient spin current inducing a Rh moment, we increase the size of the Rh moment at the beginning of each time step using a temporal Gaussian pulse
\begin{equation}
    \mathbf{S}_{i,\mathrm{Rh}}' = \mathbf{S}_{i,\mathrm{Rh}} + \frac{\Delta}{2\pi\sigma}\exp{\left(-\frac{(t-t_0)^2}{2\sigma^2}\right)\hat{\mathbf{J}}},
\end{equation}
which has a width of $\sigma$ in dimensions of time, a temporal center of $t_0$, a dimensionless amplitude of $\Delta$ and $\hat{\mathbf{J}}$ is a unit vector in the direction of the polarisation of the incoming spin current. 

Within the model we assume that the FeRh N{\'e}el vector is at 45 degrees to the [100] (along $x$ in Fig.~1(a) of the main text) direction~\cite{Xie_AIPAdv_7_056314_2017, Xie_NPGAsiaMater_12_67_2020} along which the field is applied and the ferromagnetic regions will align. Therefore, $\hat{\mathbf{e}}=(\sqrt{2}/2, \sqrt{2}/2, 0)$ and the transient spin current is polarised along $\hat{\mathbf{J}}=(1, 0, 0)$.

\begin{table}[h]
\caption{\label{tab:asd_model_params} Model parameters for the generalised Langevin spin equation \eqref{eq:gse}.}
\begin{ruledtabular}
\begin{tabular}{ldl}
    symbol   & \text{value} & units \\ \hline
    $\gamma$    & $0.17608596$ & $\mathrm{rad}\cdot\mathrm{ps}^{-1}\cdot\mathrm{T}^{-1}$ \\
    $\lambda$    & $0.1$ & (dimensionless) \\
    $\mu_{\mathrm{Fe}}$ & $3.2$\footnotemark[4]  & $\mu_\mathrm{B}$ \\
    $\mu_{\mathrm{Rh}}$ & $1.0$\footnotemark[4]  & $\mu_\mathrm{B}$ \\
\end{tabular}
\end{ruledtabular}
\footnotetext[4]{Ref.~\onlinecite{Polesya_PhysRevB_93_024423_2016}}
\end{table}

\begin{table*}[h]
\caption{\label{tab:asd_exchange} Exchange constants for the Hamiltonian~\eqref{eq:gse_hamiltonian}. A single interaction vector is given in fractional coordinates; equivalent vectors can be generated from the m$\overline{3}$m point group symmetry operations. The interaction distances are given in units of the lattice constant ($a$).}
\begin{ruledtabular}
\begin{tabular}{lccccc}
    symbol   & types & vector (fractional coordinates) & number & \text{distance (lattice constants $a_0$)} & \text{value (meV)} \\ \hline
    $J_{1}$    & Fe-Rh & $(1/2,1/2,1/2)$ & $8$  & $0.866025$ & $27.16$\footnotemark[1]  \\
    $J_{2}$    & Fe-Fe & $(\quad 0,\quad 0, \quad 1)$ & $6$  & $1.000000$ & $-3.90$\footnotemark[2]  \\
    $J_{3}$    & Fe-Fe & $(\quad 0,\quad 1, \quad 1)$ & $12$  & $1.414214$ & $5.84$\footnotemark[2]  \\
    $J_{4}$    & Fe-Rh & $(1/2,1/2,3/2)$ & $24$  & $1.658312$ & $1.54$\footnotemark[1] \\
    $J_{5}$    & Fe-Fe & $(\quad 1,\quad 1, \quad 1)$ & $8$  & $1.732051$ & $-16.34$\footnotemark[2]  \\ 
    $J_{6}$    & Fe-Fe & $(\quad 0,\quad 0, \quad 2)$ & $6$  & $2.000000$ & $2.76$\footnotemark[2]  \\
    $J_{7}$    & Fe-Fe & $(\quad 0,\quad 1, \quad 2)$ & $24$  & $2.236068$ & $0.80$\footnotemark[2]  \\
    $J_{8}$    & Fe-Fe & $(\quad 1,\quad 1, \quad 2)$ & $24$  & $2.449490$ & $-1.32$\footnotemark[2]  \\
   \colrule
    $B_{1}$    & Fe-Fe & $(\quad 0,\quad 0, \quad 1)$ & $6$  & $1.000000$ & $5.00$ \\
    \colrule
    $k$          & Fe    & - & - & - & $0.06242$ \\ 
   \colrule    
   $L_{2}$       & Fe    & - & - & - & $-440.987$\footnotemark[3] \\
   $L_{4}$       & Fe    & - & - & - & $150.546$\footnotemark[3] \\
   $L_{6}$       & Fe    & - & - & - & $50.6794$\footnotemark[3] \\
   \colrule
   $L_{2}$       & Rh    & - & - & - & $100.64$ \\
   $L_{4}$       & Rh    & - & - & - & $0.0$ \\
   $L_{6}$       & Rh    & - & - & - & $0.0$
\end{tabular}
\end{ruledtabular}
\footnotetext[1]{Ref.~\onlinecite{Polesya_PhysRevB_93_024423_2016}, Fig.~4b}
\footnotetext[2]{Ref.~\onlinecite{Polesya_PhysRevB_93_024423_2016}, Fig.~4a, `DLM' dataset}
\footnotetext[3]{Ref.~\onlinecite{Ellis_PhysRevB_100_214434_2019}}
\end{table*}

\section*{S5. Estimate of Rh exchange field}

The sudden appearance of the Rh moment couples to the Fe through a strong exchange interaction. For the Fe in the AF state to generate a net moment the system must overcome the critical field for a spin-flop transition. The exchange field exists for such a short time that the spin-flop state is only transient, but a net magnetisation is produced (Fig.4b main text – there is a net Fe magnetisation) in the AF and a corresponding spin pumping. We can put this on a more quantitative footing. The FeRh AFM state is dominated by the antiferromagnetic $J_5$ interaction, and the effective exchange field felt by the Fe sublattices has a field strength,
\begin{equation}
    B_{E} = \frac{8|J_5|}{\mu_{\mathrm{Fe}}}.
\end{equation}
The anisotropy field strength is,
\begin{equation}
    B_A = \frac{2k}{\mu_{\mathrm{Fe}}}.
\end{equation}
The lower bound for the critical field of a two sublattice antiferromagnet with easy axis (at $T=0$) is~\cite{gurevich_book},
\begin{equation}
    B_0 > \sqrt{B_A(2B_E - B_A)}
\end{equation}
where $B_0$ is a field strength applied along the easy axis. For our model of FeRh $B_0 \approx 30$~Tesla. Our applied field is much lower than this, meaning that the two AFM modes can be split only a small amount, leading to only a small spin current, proportional to the applied field as seen in normal AFs. However, the spontaneously generated Rh moment causes the sudden appearance of an additional exchange field inside of the AFM. This field is not staggered (alternating between sublattices) but directional, due to the spin polarised current. Hence, it acts on the antiferromagnetic Fe sublattices in a similar way to an applied field, but is much stronger. The field strength is approximately,
\begin{equation}
    B_{\mathrm{Rh}} = \frac{8|J_1|}{\mu_{\mathrm{Fe}}}S_{\mathrm{Rh}}.
\end{equation}
Our results show that we only generate $S_{\mathrm{Rh}} = 0.2\mu_{\mathrm{B}}/1.0\mu_{\mathrm{B}}$ (Fig. 4b main text), but nevertheless the peak Fe-Rh exchange field is therefore $B_{\mathrm{Rh}} = 235$~Tesla, much greater than the spin-flop field. This explains why the Fe develops a net magnetisation, it is trying to spin-flop, although does not have enough time to fully transition in the ~1 ps the field appears for. This also explains why the spin pumping is far larger than can be achieved from an applied field below the spin-flop field of an antiferromagnet, and more similar to the large change seen in spin pumping once the field is above the spin flop transition~\cite{Wu_PhysRevLett_116_097204_2016}. 
\newpage

\bibliography{references}